\RequirePackage{fix-cm}
\documentclass[twocolumn]{svjour3} 
\smartqed
\usepackage{fixltx2e,multirow}
\usepackage{endnotes,graphicx,fullpage,amsmath}
\renewcommand{\[}{\begin{equation}}
\renewcommand{\]}{\end{equation}}
\newcommand{\eg}{e.g., }
\newcommand{\ie}{i.e., }
\renewcommand{\vec}[1]{\mathbf{\boldsymbol{#1}}}
\newcommand{\e}[1]{\hat{\vec{#1}}}
\newcommand{\unit}[1]{\,\mathrm{#1}}
\renewcommand{\r}{R}
\newcommand{\figref}[1]{Figure~\ref{#1}}
\newcommand{\tabref}[1]{Table~\ref{#1}}
\usepackage{tikz}
\newcommand{\MD}[0]{Mindlin--Deresiewicz}

\journalname{Granular Matter}
\newcommand{\degree}{^\circ}
\newcommand{\new}[1]{{#1}}
\renewcommand{\read}[1]{{#1}}

\newcommand{\red}[1]{{#1}}
\newcommand{\redd}[1]{{#1}}
\newcommand{\h}{h}
\newcommand{\n}{\mathrm{n}}
\renewcommand{\l}{\mathrm{l}}
\renewcommand{\i}{\mathrm{i}}
\newcommand{\s}{\mathrm{s}}
\newcommand{\p}{\mathrm{p}}

\newcommand{\ro}{\mathrm{ro}}
\let\TO\to 
\renewcommand{\to}{\mathrm{to}}
\newcommand{\el}{\mathrm{el}}
\newcommand{\diss}{\mathrm{diss}}
\newcommand{\rel}{\mathrm{rel}}
\newcommand{\rail}{\mathrm{rail}}
\newcommand{\la}{\lambda}
\newcommand{\quotes}[1]{\lq\lq #1\rq\rq}
\begin{document}

\title{Rolling, sliding \& torsion of micron-sized silica particles}
\subtitle{Experimental, numerical and theoretical analysis}

\author{
Regina Fuchs${}^\dagger$ \and Thomas Weinhart${}^\dagger$ \and Jan Meyer \and Hao Zhuang \and 
\mbox{Thorsten Staedler \and Xin Jiang \and Stefan Luding}
\\{\small${}^\dagger$ Both authors contributed equally to this study}
}

\institute{
R. Fuchs \and J. Meyer \and H. Zhuang \and T. Staedler \and X. Jiang \at 
Institute of Materials Engineering, University of Siegen, Paul-Bonatz-Str. 9-11, 57076 Siegen, Germany \\\email{thorsten.staedler@uni-siegen.de}        
\vspace*{-.03in}
\and
T. Weinhart \and S. Luding \at
Multiscale Mechanics, CTW, MESA+, Univ. Twente, P.O. Box 217, 7500 AE Enschede, The Netherlands
\\\email{t.weinhart@utwente.nl}        
}

\date{ }

\maketitle

\begin{abstract}
%
%
The contact mechanics of individual, very small particles with other particles and walls is studied using a nanoindenter setup that allows normal and lateral displacement control and measurement of the respective forces. The sliding, rolling and torsional forces and torques are tested with borosilicate microspheres, featuring radii of about $10\,\mathrm{\mu m}$. The contacts are with flat silicon substrates of different roughness for pure sliding and rolling and with silicon based, ion-beam crafted rail systems for combined rolling and torsion.

The experimental results are discussed and compared to various analytical predictions and contact models, allowing for two concurrent interpretations of the effects of surface roughness, plasticity and adhesion. This enables us to determine both rolling and torsion friction coefficients together with their associated length scales. Interestingly, even though normal contacts behave elastically \linebreak[4](Hertzian), all other modes of motion display effects due to surface roughness and consequent plastic deformation.
The influence of adhesion is interpreted in the framework of different models and is very different for different degrees of freedom, being largest for rolling.

\vspace*{-.07in}
\end{abstract}

\newpage
\section{Introduction} \label{sec:intro}
\vspace*{-.1in}
A plethora of applications in pharmacy, cosmetics, food industry and other areas are directly linked to the research field of particle technology and contact mechanics in particular. The latter is a challenging topic as the characteristics of individual particle contacts and details of the contacting surfaces determine the behavior of whole ensembles. Consequently any modeling approach requires appropriate experimental data on the corresponding particle and, even smaller, on the contact scale.

In general, experimental techniques can be classified as being either contact or non-contact based. In the following we will focus on the existing work in context of contact based techniques. These feature the biggest potential in probing individual particle contacts. Here, the advent of scanning pro\-be microscopy paved the way of today's most prominent contact method to determine particle--substra\-te adhesion\,\cite{Ducker1991,Butt1991} as well as particle motions\, \cite{Sitti2000,Sitti2004,Sumer2008}. The atomic force microscope (AFM) colloid probe technique opens up the possibility of quantitative particle--substrate and particle--par\-ti\-cle adhesion measurements in a low load regime. Especially for particles, which are not perfect spheres, one of the key features affecting their contacts is surface roughness. An increasing roughness tends to increase the mean separation between two interacting bodies, which results in a decrease of adhesion. Liu et al.\,\cite{Liu2007} used the colloid probe technique to show the existence of an optimal amount of asperity yielding a minimum adhesion. This behavior was also reported by Rabinovich et al.\,\cite{Rabinovich2000a,Rabinovich2000b} and Matope et al.\,\cite{Matope2012}. The direct effect of surface roughness on particle motion was simulated by Karayem and Zakeri\,\cite{Korayem2011}, who reported that the critical force for particle motion decreases when the particles are pushed on rough surfaces for rolling as well as sliding motion. Furthermore, qualitative results showed that for particle rolling an external moment has to overcome a certain threshold\,\cite{Saito2002}, also known as critical moment or rolling resistance moment\,\cite{Peri2008,Ding2007} to initiate a rolling motion. \new{All forecited work 
used a combination of lateral and normal load to initiate rolling, \eg by using an AFM cantilever as a pushing tool. Rolling initialization with a controlled normal force was not realized until the work of Vilt et al.\,\cite{Vilt2011}. They} presented a fundamental study of frictional properties of silica microsphere monolayers on silicon substrates by using ball-on-flat tribometer and reported that rolling resistance is highly dependent on load and sphere diameter. 

However, all of the above mentioned studies either require dedicated self-made set-ups or are limited with respect to the weight of the particle and/or the accessible load regime. With respect to these limits, nanoindentation appears to be a suitable alternative. Additionally, recent technological advances allow nanoindentation to be carried out in a load regime previously reserved to AFM settings. 

In this work, the idea of the colloid probe technique is transferred to a nanoindenter setup which allows for the preparation of larger particle probes, higher maximum normal loads, and a unique strategy to sample rolling and torsion friction of individual particles. This tool is then used to study the sliding, rolling and torsion friction acting on individual micron-sized borosilicate spheres in contact with Si surfaces. The experimental techniques are detailed in chapter \ref{sec:expd}.

To understand, classify and describe the particle motion from the modeling side, a closer look at the various acting forces and torques was done \new{in chapter \ref{sec:cm}}. The forces acting at a contact can be due to many different physical effects; here, we will take adhesive, elastic and dissipative normal forces and frictional tangential forces into account. 

Elastic deformations of spheres are described by a Hertz\-ian normal force\,\cite{hertz1882b}. These models can be extended to model more complex effects such as elasto-plastic deformation\,\cite{Luding2008c,Tomas20075925,SinghMagnanimoLuding2013}, or various other interaction mechanics -- however, here we restrict ourselves to elastic Hertz contacts since they are valid in most of the experiments for the materials, sizes and compression forces described below in chapter \ref{sec:expd}. \new{The validity of the Hertzian contact model for silica nanoparticles was recently confirmed by molecular dynamics simulations\,\cite{SunZengYu2013}}.

Three friction-like mechanisms can be distinguished that act against different relative motions other than normal displacement; sliding, rolling and torsion friction. \new{They are quantified by the sliding friction coefficient, which is the ratio between the tangential and the normal force, and the rolling and torsion friction factors, which are the ratios between the rolling and torsion torque, respectively, and the normal force.}
\read{Rolling and torsion friction factors have units of length, and can thus be split into a dimensionless friction coefficient and a length scale.}
If the friction coefficient is constant, it can be called Coulomb-type. 
Experimentally observed contact forces can be used to test and calibrate contact force models for the simulation of granular materials. 
\new{However, the contact forces can rarely be derived theoretically from the material and surface properties, or measured directly in experiments, as particles are often very small, difficult to handle, and particle shapes and sizes can be very diverse, \ie different from perfectly spherical.}

\new{
We further show how to decompose the tangential forces into sliding, rolling and torsion \quotes{friction} \cite{Luding2008c}, and discuss various contact models. Each of these forces as well as the different torques play an important part in the bulk behavior and cannot be ignored in many cases; in particular, rolling friction is necessary to model the roughness (and thus rolling resistance) of macroscopically smooth surfaces \cite{ShojaaeeBrendelTorokWolf2012} and affects the shear behavior \cite{WangZhuLudingYu2013}.

The experimental results in chapter \ref{sec:expr} are linked to different contact models in order to determine an appropriate one that reproduces the experimental findings. 
From the model we show how to obtain some of the corresponding sliding, rolling and torsion friction coefficients.
Care is taken to account for the influence of surface roughness and adhesion on the frictional properties. 
}
Finally, the analytical results are compared with particle simulations, so that their implementation in a many-particle simulation environment is only a tiny further step.  

\vspace*{-.1in}
\section{Experimental details} \label{sec:expd}

Prior to the introduction of the nanoindentation based testing techniques for characterizing the sliding, rolling and torsion friction of individual particles, see Figures~\ref{fig:sliding}-\ref{fig:sem_rail}, details of the spherical probe particles as well as the corresponding Si surfaces will be presented. Additionally, a brief description of the procedures, that have been utilized to acquire the root mean squared (RMS) roughness of the silicon surfaces as well as the adhesion between particles and those surfaces, will be given.

\subsection{Test objects}

\subsubsection{Surfaces and spherical particles}

The substrate material used was a single-crystalline Si(100) wafer, supplied by Siegert Wafer GmbH, Aachen. Samples were cleaned by exposure to chloroform (CHCl\textsubscript{3}, 99\%, Roth), afterward by rinsing with ethanol (EtOH, 97\% with 1\% petroleumether, Roth) and Milli-Q water, followed by drying in a stream of nitrogen.

Surface topography changes of these samples were achieved through a slow etching process using H\textsubscript{2} plasma. The modification is carried out in an ASTeX A5000 microwave plasma chemical vapor deposition (MWCVD) reactor with an H\textsubscript{2} flow rate of \new{400 standard cubic centimeters per minute}. Two different microwave powers (1600\,W and 1800\,W coupled with a gas pressure of 40\,Torr and 45\,Torr, respectively) are used to sustain the plasma. The etching period was 20 minutes for all samples. This treatment leads to a variation in surface roughness due to the different etching efficiency of the H\textsubscript{2} plasma with different microwave power. Afterwards, the modified samples were stored under ambient conditions for two weeks, allowing the formation of a thin natural oxide layer (${\sim}1\,$nm, see \cite{morita1989control}).

A Focused-Ion beam (FIB) system (FEI Helios 600) was used to create rail-structure silicon samples (see \figref{fig:sem_rail}) featuring a length of  $100\,\mathrm{\mu m}$ and rail inclinations of $25\degree$, $45\degree$ and $60\degree$, respectively. Depth and width of the rails were chosen in such a way that the center of mass of the particles was situated $1\unit{\mu{}m}$ below the wafer's surface level. This design effectively keeps the particles inside the rail during an experiment. Scanning electron microscopy (Field Emission Scanning Electron Microscope, FESEM, Ultra 55, Zeiss) was used for rail angle characterization (see \figref{fig:sem_rail} below).

Regarding model particles, guided by the work of van Zwol et al.\,\cite{vanZwol2008} we decided in favor of borosilicate glass beads provided by Duke Scientific that feature nominal particle radii of $10\,\mathrm{\mu m}$ (Duke Standards 9020, mean diameter $17.3\,\mathrm{\mu m} \pm 1.0\,\mathrm{\mu m}$, size distribution $2.0\,\mathrm{\mu m}$ std. dev.). The RMS roughness of these micro-spheres is $0.7 \pm 0.1\,$nm \,\cite{Liu2007}. 

\subsubsection{Colloid and other probes}

AFM colloidal probes were prepared by attaching (Araldite 10\,min, 2 components, Epoxy) the borosilicate glass spheres to Mikromasch NSC12 tip-less AlBS cantilevers. The radii of the colloid probes and the quality (clean contact area) were determined by using scanning electron microscopy.

In order to prepare colloid probes for the scanning nanoindenter (TriboIndenter, Hysitron Inc.) used in this work, the FIB was employed to create appropriate cavities into the apex of a commercially available diamond cube corner tip (Hysitron Inc.), see \figref{fig:sliding} below. Subsequently, the borosilicate spheres were fixed to this holder by means of a small amount of photosensitive acrylate-based adhesive glue (DIC Europe GmbH). The quality of each of the probes created in this manner was evaluated by scanning electron microscopy as well as a series of indents with varying maximal load, ranging from 1 to 10\,mN, into a smooth fused silica surface and sequential fitting of the resulting load-displacement curves by assuming a Hertzian contact. Any probe which resulted in identical estimates of both characterization techniques for the sphere radius was employed for further tribological testing like normal and sliding tests.

A flat end diamond indenter ($20\,\mathrm{\mu m}$ diameter, Hysitron Inc.) and non-glued spheres were utilized to sample rolling and torsional friction, as detailed in \cite{FuchsMeyerStaedlerJiang2013}. The misalignment of the indenter axis was below $1\degree$ and the RMS roughness of the central part of the punch was measured to be 4$\pm$\,1 nm (lateral scale of $2.5\,\mathrm{\mu m}$), see \figref{fig:rolling_sem+schema}. 

\subsection {Roughness and adhesion measurements}

Atomic force microscopy (XE-100, PSIA) equipped with commercially available tips (ACTA, AppNano) featuring tip radii below $10\,$ nm in non-contact mode was used to obtain information about the surface roughness. 

The untreated Si(100) wafers, the etched Si surfaces as well as the rail-structured Si surfaces were characterized each at five different spots. Scan-area and pixel resolution were chosen to be $2.5\times2.5\,\mathrm{\mu m^{2}}$ and $512\!\times\!512$, respectively. The RMS roughness, at each of the five different spots, was calculated after a simple linear tilt correction in \textit{x}- and \textit{y}-direction. The RMS roughness was then defined as the mean RMS roughness of the five individual spots. 

AFM (Asylum Research MFP-3DTM AFM, Santa Barbara, CA) based adhesion measurements were carried out with the colloid probes. The colloid probe cantilever spring constant was evaluated by utilizing the thermal noise method \cite{ButtJaschke1995}. Force-dis\-tance curves were measured using the force volume (force map) method, which automatically probes $20\times20$ points in a $50\times50\,\mathrm{\mu m^{2}}$ area. The adhesion forces were sampled using a $z$-rate of $1.0\,\mathrm{\mu m/s}$. The loading force used was \textless 5-10\% of the adhesion forces. All experiments were conducted at room temperature and 33\% relative humidity. The final adhesion force of a given sample was calculated as the mean value of 400 individual measurements. 

\subsection {Sliding tests}

\begin{figure}[t] 
	\centering
		\includegraphics[width=.5\columnwidth]{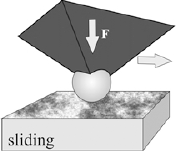}
		\caption{Schematic of the test setup for the sliding experiments of individual fixed micron-sized particles.}
	\label{fig:sliding}
\end{figure}

In order to characterize the sliding friction of the particles in the predominant elastic contact regime, nanoindenter colloid probe tips have been used according to the schematic given in \figref{fig:sliding}.

The sliding friction tests were performed in a nano\-indentation based setup using a TriboIndenter in combination with a Performech upgrade (Hysitron Inc.). The actual tests to characterize sliding friction were carried out at room temperature and $30\pm 5\%$ relative humidity in a friction loop fashion, which is a common technique in the AFM community. In this case the probes have been scratched back and forth over a distance of $2\,\mathrm{\mu m}$ at a constant normal load and a speed of $1\,\mathrm{\mu m/s}$ under load control. The normal load was varied between 3 and $100\,\mathrm{\mu N}$; five spheres were utilized probing each load five times at different surface spots for all Si surfaces, respectively. In turn, the corresponding absolute value of lateral force for each test was evaluated by a lateral displacement sensitive averaging of the difference in measured lateral force for forward and backward movement divided by two. In order to avoid artifacts originating from a change in the movement direction, only the central micron range of the friction loop was taken into account.

\subsection {Rolling tests}\label{sec:24}

\begin{figure}[t]
	\centering
	\begin{minipage}[b]{3.5 cm}
		\includegraphics[width=1\columnwidth]{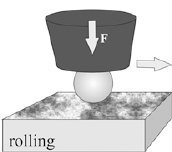}
		 \end{minipage}
  \begin{minipage}[b]{4.0 cm}
	\includegraphics[width=1\columnwidth]{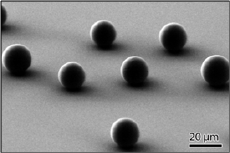}
	  \end{minipage}
		\caption{Schematic of the test setup for rolling experiments of individual micron-sized particles and FESEM image of borosilicate glass microspheres featuring a diameter of about $17\,\mathrm{\mu m}$ on a Si substrate. }
	\label{fig:rolling_sem+schema}
\end{figure}

The nanoindentation test setup for the measurement of rolling friction of the microspheres is significantly easier compared to the sliding case (\figref{fig:rolling_sem+schema}). Here, the borosilicate spheres featuring radii of approximately $10\,\mathrm{\mu m}$ are placed on the surface of interest prior to testing. In a second step individual beads are contacted with a flat end diamond indenter, carefully positioning the indenter tip with its axis aligned with respect to the apex of the spherical particle of choice. 

Preliminary test results with single scratch length of about $10\,\mathrm{\mu m}$ and a speed of $1\,\mathrm{\mu m/s}$ under load control, showed that a certain threshold\,\cite{Saito2002} or rolling resistance moment\,\cite{Peri2008,Ding2007} was necessary to initiate particle rolling. In our case, the rolling of particles, i.e. a change in position detected by optical microscopy, was only observed for normal loads larger than $100\,\mathrm{\mu N}$.

Based on these results, a minimum normal load of $100\,\mathrm{\mu N}$ was selected for all rolling tests presented here to ensure rolling characteristics of the corresponding contact behavior. This $100\,\mathrm{\mu N}$ force to overcome the rolling resistant moment, was chosen independently of the load regime used in sliding tests. 
%
The final rolling tests were carried out under the same environmental conditions in analogy to the friction loop method presented above and with normal load ranging from $100\,\mathrm{\mu N}$ to 3\,mN.

Again, in order to improve the reliability of the results, five beads were used for testing and each measurement, i.e. applied normal load, was taken five times for all Si surfaces, respectively. The absolute values of lateral forces for each test were calculated in analogy to the friction loop method presented for sliding friction tests.

\subsection{Combined rolling and torsion tests}

In combination with appropriate rail systems, the strategy to probe rolling friction, which has been introduced above, can also be exploited to study a combination of rolling and torsional friction. As far as the authors are aware, no scanning probe based technique has been proposed  so far that would allow for the exclusive evaluation of torsion friction. 

In order to access the combination of both mechanisms, we utilized Si based rail structures featuring rail angles of $25\degree$, $45\degree$ and $60\degree$. Higher rail angles lead to an increased influence of the torsional contact mode. 

\begin{figure}[t]
	\centering
		\includegraphics[width=.5\columnwidth]{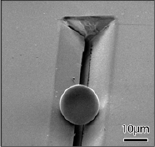}
		\caption{FESEM image of a borosilicate glass microsphere in a rail structured Si surface. The rail system was prepared by FIB. Sphere positioning was realized by AFM.}
	\label{fig:sem_rail}
\end{figure}

Borosilicate spheres are placed inside the rails by AFM based positioning (\figref{fig:sem_rail}). In analogy to rolling tests, the individual particles inside the rails are contacted and probed with a flat end diamond indenter. Careful positioning of the indenter tip with respect to the particle as well as alignment of the rail with regard to the scratch axis (movement direction) of the instrument are crucial.

It should be mentioned, that the actual normal loads acting on the rail slopes are calculated by vector considerations. 
The effect of scratch velocity can be considered negligible, since the results of the rolling test did not change while varying velocity from 1, 0.1 to $0.01\,\mathrm{\mu m/s}$  inside the rail system.

\section{\new{Contact mechanics}}\label{sec:cm}

\new{In section \ref{sec:cmech}, basic contact mechanics is applied to the system presented here. Then, in section \ref{sec:cmod}, we review the most common contact models. These models will then be compared to the experimental results in chapter \ref{sec:expr} to deduce important characteristics of the particle-substrate and particle-indenter contacts.}

\subsection{\new{Description of the contact mechanics}} \label{sec:ass} \label{sec:cmech}

\new{We will describe the contact mechanics in the context of soft sphere contacts, as is used in Discrete Element Method (DEM) simulations}, where the contact deformation is not modeled explicitly; instead it is assumed that the particles slightly overlap, with a single contact point at the center of the overlap, where a contact force and torque can be applied. \new{In the following, we will define the notation and describe a universally valid decomposition of the contact forces and velocities}. 

\subsubsection{Particles and surfaces}

Both substrate and indenter surfaces, are assumed to be planar, while the particle is assumed to be (perfectly) spherical.
\new{To describe the particle-plane contact}, we denoted the radius of the particle as $\r$, mass $m$, position $\vec{r}_\p$, velocity $\vec{v}_\p$ and angular velocity around the center of mass $\vec{\omega}_\p$.
%
%
The contact point between particle and surface is denoted by $\vec{c}$,  the relative distance vector $\vec{r}=\vec{r}_\p-\vec{c}$, the separation, or corrected radius, $r=|\vec{r}|$, and the unit normal $\hat{\vec{n}} = \vec{r}/r$.
\read{Subscripts $\i$, $\s$ are used where necessary to distinguish between the particle contact with the indenter and the substrate, respectively. Therefore, we denoted the relative velocity of the surfaces at the contact with $\vec{v}^{\rel}$, with $\vec{v}_\i^{\rel} = \vec{v}_\p - \vec{v}_\i + \vec{r} \times \vec{\omega}_\p$ for the indenter-particle contact and $\vec{v}_\s^{\rel} = \vec{v}_\p + \vec{r} \times \vec{\omega}_\p$ for the substrate-particle contact (the substrate does not move). The relative angular velocity is always  $\vec{\omega}^{\rel}=\vec{\omega}_\p$ since the substrate and indenter do not rotate.} For a schematic, see \figref{fig:forces0} below.

\begin{figure*}[t]
\begin{minipage}{.32\textwidth}
\scalebox{1}{\begin{tikzpicture}[scale=0.33]
\def\forcecolor{blue}
\def\torquecolor{green}
\def\N{\mathrm{n}}
\def\L{\mathrm{l}}
\def\S{\mathrm{s}}
\def\I{\mathrm{i}}

\def\vectorout{\fill[color=white] (\x,\y) circle(.28);\draw (\x,\y) circle(.28); \fill[color=black] (\x,\y) circle(.05);}
\def\vectorin{\fill[color=white] (\x,\y) circle(.28);\draw (\x,\y) circle(.28);\draw (\x-.21,\y-.21) --(\x+.21,\y+.21); \draw (\x+.21,\y-.21) --(\x-.21,\y+.21);}
\def\forceout{\fill[\forcecolor,color=white] (\x,\y) circle(.2);\draw[\forcecolor] (\x,\y) circle(.2); \fill[color=\forcecolor] (\x,\y) circle(.05);}
\def\forcein{\fill[\forcecolor,color=white] (\x,\y) circle(.2);\draw[\forcecolor] (\x,\y) circle(.2);\draw[\forcecolor] (\x-.14,\y-.14) --(\x+.14,\y+.14); \draw[\forcecolor] (\x+.14,\y-.14) --(\x-.14,\y+.14);}
\def\torqueout{\fill[color=white,\torquecolor] (\x,\y) circle(.2);\draw[\torquecolor] (\x,\y) circle(.2); \fill[color=\torquecolor] (\x,\y) circle(.05);}
\def\torquein{\fill[color=white,\torquecolor] (\x,\y) circle(.2);\draw[\torquecolor] (\x,\y) circle(.2);\draw[\torquecolor] (\x-.14,\y-.14) --(\x+.14,\y+.14); \draw[\torquecolor] (\x+.14,\y-.14) --(\x-.14,\y+.14);}

\def\ozyx{\draw[->] (\x,\y) --(\x+\l,\y); \node[below] at (\x+\l,\y) {$\hat{\vec z}$};
\draw[->] (\x,\y) --(\x,\y+\l); \node[right] at (\x,\y+\l) {$\hat{\vec y}$};
\vectorin \node[below left] at (\x,\y) {$\hat{\vec x}$};}

\def\x{20}\def\y{3}\def\l{2} \ozyx

\draw (21,10) --(34,10); 
\filldraw[black,fill=white!20!white] (27.5,5) circle(5);
\filldraw[black,fill=white!20!white](25,9.33) -- (25,10) --(30,10) --(30,9.33); 
\draw (21,0) --(34,0); 

\def\x{21}\def\y{10.3} \vectorin \node[above] at (\x,\y) {$\vec v_i$};
\def\x{27.5}\def\y{5} \vectorin \node[below] at (\x,\y) {$\vec v$}; 

\def\x{27.5}\def\y{9.9}

\def\x{27.5}\def\y{-0.1}
\def\s{0}\def\c{2.5}
\draw[\forcecolor,->] (\x,\y) --(\x+\s,\y+\c); \node[\forcecolor,right] at (\x+\s,\y+\c) {$\vec f^{\N}_{\S}$};
\forceout \node[\forcecolor,below] at (\x,\y) {$\vec f^{\L}_{\S}$}; 
\end{tikzpicture}}
\centering\\(a)
\end{minipage}\hfill
\begin{minipage}{.32\textwidth}
\scalebox{1}{\begin{tikzpicture}[scale=0.33]
\def\vectorout{\fill[color=white] (\x,\y) circle(.28);\draw (\x,\y) circle(.28); \fill[color=black] (\x,\y) circle(.05);}
\def\vectorin{\fill[color=white] (\x,\y) circle(.28);\draw (\x,\y) circle(.28);\draw (\x-.21,\y-.21) --(\x+.21,\y+.21); \draw (\x+.21,\y-.21) --(\x-.21,\y+.21);}
\def\forcecolor{blue}
\def\torquecolor{green!50!red}
\def\N{\mathrm{n}}
\def\L{\mathrm{l}}
\def\S{\mathrm{s}}
\def\I{\mathrm{i}}
\def\ro{\mathrm{ro}}

\def\forceout{\fill[\forcecolor,color=white] (\x,\y) circle(.28);\draw[\forcecolor] (\x,\y) circle(.28); \fill[color=\forcecolor] (\x,\y) circle(.05);}
\def\forcein{\fill[\forcecolor,color=white] (\x,\y) circle(.28);\draw[\forcecolor] (\x,\y) circle(.28);\draw[\forcecolor] (\x-.21,\y-.21) --(\x+.21,\y+.21); \draw (\x+.21,\y-.21) --(\x-.21,\y+.21);}
\def\torqueout{\fill[color=white,\torquecolor] (\x,\y) circle(.28);\draw (\x,\y) circle(.28); \fill[color=\torquecolor] (\x,\y) circle(.05);}
\def\torquein{\fill[color=white,\torquecolor] (\x,\y) circle(.28);\draw (\x,\y) circle(.28);\draw (\x-.21,\y-.21) --(\x+.21,\y+.21); \draw (\x+.21,\y-.21) --(\x-.21,\y+.21);}

\def\oxyz{\draw[->] (\x,\y) --(\x+\l,\y); \node[right] at (\x+\l,\y) {$\vec x$};
\draw[->] (\x,\y) --(\x,\y+\l); \node[right] at (\x,\y+\l) {$\vec y$};
\vectorout \node[below left] at (\x,\y) {$\vec z$};}
\def\ozyx{\draw[->] (\x,\y) --(\x+\l,\y); \node[right] at (\x+\l,\y) {$\vec z$};
\draw[->] (\x,\y) --(\x,\y+\l); \node[right] at (\x,\y+\l) {$\vec y$};
\vectorin \node[below left] at (\x,\y) {$\vec x$};}


\draw (21,10) --(34,10); 
\draw (27.5,5) circle(5);
\draw (21,0) --(34,0); 

\draw[->] (27.5,5) --(25,5); \node[above] at (25,5) {$\vec \omega$};
\def\x{21}\def\y{10.3} \vectorin \node[above] at (\x,\y) {$\vec v_{\I}$};
\def\x{27.5}\def\y{5} \vectorin \node[below] at (\x,\y) {$\vec v$}; 

\def\x{27.5}\def\y{9.9}
\draw[\torquecolor,->,dashed] (\x,\y) --(\x+2.5,\y); \node[\torquecolor,above] at (\x+2.5,\y) {$\vec M^{\ro}_{\I}$};
\draw[\torquecolor,->,dashed] (\x,\y) --(\x-2.5,\y); \node[\torquecolor,above] at (\x-2.5,\y) {$\vec M^{\L}_{\I}$};
\draw[\forcecolor,->] (\x,\y) --(\x,\y-2.5); \node[\forcecolor,right] at (\x,\y-2.5) {$\vec f^{\N}_{\I}$};
\forcein \node[\forcecolor,above] at (\x,\y) {$\vec f^{\L}_{\I}$}; 

\def\x{27.5}\def\y{-0.1}
\def\s{0}\def\c{2.5}
\draw[\torquecolor,->,dashed] (\x,\y) --(\x+\c,\y+\s); \node[\torquecolor,below] at (\x+\c,\y+\s) {$\vec M^{\ro}_{\S}$};
\draw[\torquecolor,->,dashed] (\x,\y) --(\x-\c,\y-\s); \node[\torquecolor,below] at (\x-\c,\y-\s) {$\vec M^{\L}_{\S}$};
\draw[\forcecolor,->] (\x,\y) --(\x+\s,\y+\c); \node[\forcecolor,right] at (\x+\s,\y+\c) {$\vec f^{\N}_{\S}$};
\forceout \node[\forcecolor,below] at (\x,\y) {$\vec f^{\L}_{\S}$}; 
\end{tikzpicture}}
\centering\\(b)
\end{minipage}\hfill
\begin{minipage}{.32\textwidth}
\scalebox{1}{\begin{tikzpicture}[scale=0.33]
\def\vectorout{\fill[color=white] (\x,\y) circle(.28);\draw (\x,\y) circle(.28); \fill[color=black] (\x,\y) circle(.05);}
\def\vectorin{\fill[color=white] (\x,\y) circle(.28);\draw (\x,\y) circle(.28);\draw (\x-.21,\y-.21) --(\x+.21,\y+.21); \draw (\x+.21,\y-.21) --(\x-.21,\y+.21);}
\def\forcecolor{blue}
\def\torquecolor{green!50!red}
\def\N{\mathrm{n}}
\def\L{\mathrm{l}}
\def\S{\mathrm{s}}
\def\I{\mathrm{i}}
\def\ro{\mathrm{ro}}
\def\to{\mathrm{to}}

\def\forceout{\fill[\forcecolor,color=white] (\x,\y) circle(.28);\draw[\forcecolor] (\x,\y) circle(.28); \fill[color=\forcecolor] (\x,\y) circle(.05);}
\def\forcein{\fill[\forcecolor,color=white] (\x,\y) circle(.28);\draw[\forcecolor] (\x,\y) circle(.28);\draw[\forcecolor] (\x-.21,\y-.21) --(\x+.21,\y+.21); \draw (\x+.21,\y-.21) --(\x-.21,\y+.21);}
\def\torqueout{\fill[color=white,\torquecolor] (\x,\y) circle(.28);\draw (\x,\y) circle(.28); \fill[color=\torquecolor] (\x,\y) circle(.05);}
\def\torquein{\fill[color=white,\torquecolor] (\x,\y) circle(.28);\draw (\x,\y) circle(.28);\draw (\x-.21,\y-.21) --(\x+.21,\y+.21); \draw (\x+.21,\y-.21) --(\x-.21,\y+.21);}



\draw (21,10) --(34,10); 
\draw (27.5,5) circle(5);
\def\x{27.5}\def\y{-1}
\draw (\x,\y) --(\x+6.5,\y+4.2);
\draw (\x,\y) --(\x-6.5,\y+4.2);

\draw (\x,\y) --(\x+2.5,\y); \node[above right] at (\x+1.5,\y) {$\theta$};
\draw (\x+2.5,\y) arc (0:33:2.5);

\draw[->] (27.5,5) --(25,5); \node[above] at (25,5) {$\vec \omega$};
\def\x{21}\def\y{10.3} \vectorin \node[above] at (\x,\y) {$\vec v_{\I}$};
\def\x{27.5}\def\y{5} \vectorin \node[right] at (\x,\y) {$\vec v$}; 

\def\x{27.5}\def\y{9.9}
\draw[\torquecolor,->,dashed] (\x,\y) --(\x+2.5,\y); \node[\torquecolor,above] at (\x+2.5,\y) {$\vec M^{\ro}_{\I}$};
\draw[\torquecolor,->,dashed] (\x,\y) --(\x-2.5,\y); \node[\torquecolor,above] at (\x-2.5,\y) {$\vec M^{\L}_{\I}$};
\draw[\forcecolor,->] (\x,\y) --(\x,\y-2.5); \node[\forcecolor,right] at (\x,\y-2.5) {$\vec f^{\N}_{\I}$};
\forcein \node[\forcecolor,above] at (\x,\y) {$\vec f^{\L}_{\I}$}; 

\def\s{1.35678846449}\def\c{2.09979167124}
\def\x{27.5+2.05*\s}\def\y{5-2.05*\c}
\draw[\torquecolor,->,dashed] (\x,\y) --(\x+\c,\y+\s); \node[\torquecolor,below right] at (\x+\c,\y+\s) {$\vec M^{\ro}_{\S'}$};
\draw[\torquecolor,->,dashed] (\x,\y) --(\x-\c,\y-\s); \node[\torquecolor,above] at (\x-\c,\y-.6*\s) {$\vec M^{\L}_{\S'}$};
\draw[\torquecolor,->,dashed] (\x,\y) --(\x+\s,\y-\c); \node[\torquecolor,above right] at (\x+\s,\y-\c) {$\vec M^{\to}_{\S'}$};
\draw[\forcecolor,->] (\x,\y) --(\x-\s,\y+\c); \node[\forcecolor,above right] at (\x-\s,\y+\c) {$\vec f^{\N}_{\S'}$};
\forceout \node[\forcecolor,right] at (\x,\y) {$\vec f^{\L}_{\S'}$}; 

\def\s{1.35678846449}\def\c{2.09979167124}
\def\x{27.5-2.05*\s}\def\y{5-2.05*\c}
\draw[\torquecolor,->,dashed] (\x,\y) --(\x+\c,\y-\s); \node[\torquecolor,below left] at (\x+\c,\y-\s) {$\vec M^{\ro}_{\S}$};
\draw[\torquecolor,->,dashed] (\x,\y) --(\x-\c,\y+\s); \node[\torquecolor,below left] at (\x-\c,\y+\s) {$\vec M^{\L}_{\S}$};
\draw[\forcecolor,->] (\x+0.05*\c,\y-0.05*\s) --(\x+\s+0.05*\c,\y+\c-0.05*\s); \node[\forcecolor,above right] at (\x+\s,\y+\c) {$\vec f^{\N}_{\S}$};
\draw[\torquecolor,->,dashed] (\x-0.05*\c,\y+0.05*\s) --(\x+\s-0.05*\c,\y+\c+0.05*\s); \node[\torquecolor,above left] at (\x+\s,\y+\c) {$\vec M^{\to}_{\S}$};
\forceout \node[\forcecolor,below left] at (\x,\y) {$\vec f^{\L}_{\S}$};

\end{tikzpicture}
}
\centering\\(c)
\end{minipage}
\caption{
\label{fig:forces0}
Sketch of forces (blue arrows) and torques (yellow dashed arrows) acting on a single particle, with the indenter (top wall) moving steadily at velocity $\vec{v}_{\i}$ (in the direction of view), while the substrate (bottom wall) is fixed. In the sliding case (\emph{a}), the particle is fixed to the indenter ($\vec{v}=\vec{v}_{\i}$).
In the rolling and rail cases (\emph{b,c}), the particle is free to move ($\vec{v}\neq\vec{v}_{\i}$).
In the latter, rail case (\emph{c}), the substrate forms a v-shaped rail of inclination $\theta$.}
\end{figure*}
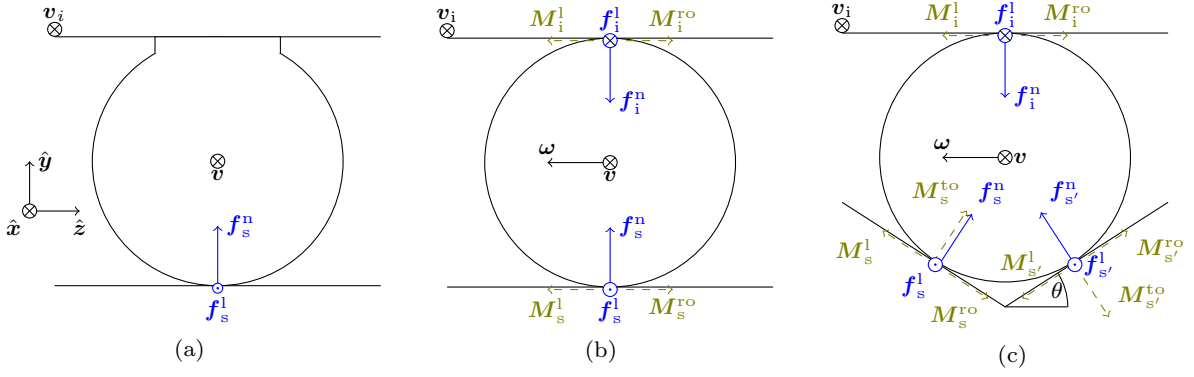

\subsubsection{Contact forces and torques}

\new{
While we defined a specific single point as the \emph{contact point} at which the contact forces and torques act, it is important to remember that they are derived from integrating the stress on the contact surface. 
Given a normal vector, the two integrals over the normal and tangential surface stress yield the normal and tangential components of the contact force,
\begin{subequations}
\begin{equation}
\vec{f} = \vec{f}^{\n} + \vec{f}^{\l}.
\end{equation}

When integrating the torque contributions (with the particle's center of mass as reference point), the torque can be decomposed into 
(\emph{i}) a torque \new{due to the tangential force acting at the contact point}, $\vec{r} \times \vec{f}^{\l}$ 
and (\emph{ii}) a remaining part, which can be further decomposed into a normal, $\vec{M}^{\to}$, and a tangential, $\vec{M}^{\ro}$, component, yielding
\begin{equation}
\vec{M} = \vec{r} \times \vec{f}^{\l} + \vec{M}^{\ro} + \vec{M}^{\to}.
\end{equation}
\end{subequations}
}

\subsubsection{Contact velocities and displacements}

The contact forces and torques acting on the particle at each contact are expected to be objective and thus independent of  the absolute (translational and angular) velocities and displacements of the observer (the frame of reference). 
\new{We reviewed and use the model by Luding\,\cite{Luding2008c} and decompose the relative (translational and angular) velocities and displacements into four  components:}

(a) 
The normal relative velocity, 
\[v^{\rel,\n}=\vec{v}^{\rel} \cdot \hat{\vec{n}} = \frac{d\delta^\n}{dt},\]
with the normal displacement, or overlap,
\[\delta^{\n}=\max(0,\r-r).\]

\begin{subequations}
(b) the lateral sliding velocity, measuring the tangential surface velocity at the particle contact,
\[ \vec{v}^{\l} = \vec{v}^{\rel} - v^{\rel,\n} \hat{\vec{n}}, \]
(c) the rolling \quotes{velocity}, measuring the rate at which two surface roll over each other,
\[ \vec{v}^{\ro} = {r} (\vec{\omega}^{\rel} \times \hat{\vec{n}}),\mbox{ and } \]
(d) the torsion \quotes{velocity}, measuring the normal relati\-ve angular velocity scaled by the effective radius,\!\!
\footnote{For analogy to the rolling velocity, we use $r$ here, even though the contact radius $a$ should be used; this definition (see below), however, only affects the torsion stiffness coefficient and has no influence on the torque.}\vspace{-.3cm}
\[ \vec{v}^{\to} = {r} (\vec{\omega}^{\rel} \cdot \hat{\vec{n}}) \hat{\vec{n}}.\]
\end{subequations}
Each of these velocities is objective, meaning that the observer will measure the same velocities, even if he is relocated by a finite or incremental translation and/or rotation. See Refs.\,\cite{Luding2008c,kuhn2004contact} for the general definition and objectiveness of the rolling velocity. 

The (elastic) displacements based on these velocities are set to zero at the initial time of contact and their rate of change is given by
\[\label{eq:deltasl}
\frac{d \vec{\delta}^{\nu}}{dt} =
\vec{v}^{\nu},\quad \nu=\l,\ro,\to.
\]
\new{Here,} the normal vector $\hat{\vec{n}}$ is constant over time; for collisions in a general reference frame, the displacements are calculated such that the directions of $\vec{\delta}^{\l}$ and $\vec{\delta}^{\ro}$ stay perpendicular, while $\vec{\delta}^{\to}$ stays parallel to the normal vector; see\,\cite{Luding2008c} for details.

\subsection{Different contact models} \label{sec:cmod}

Next, we review some contact models which satisfy the assumptions stated in chapter \ref{sec:ass}.
 
\subsubsection{Normal forces} \label{sec:nf}

\begin{figure}[btp]
\centering
\includegraphics[width=1\columnwidth]{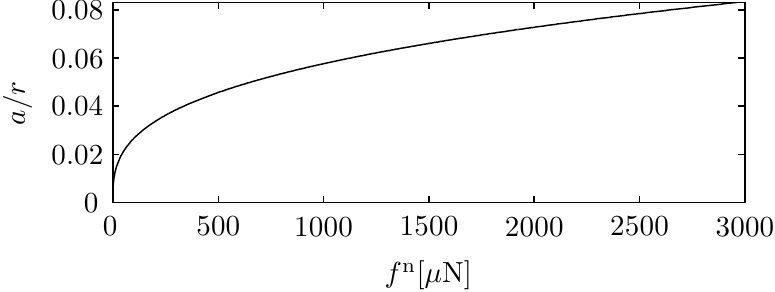}
\caption{Ratio of contact and particle radius as a function of the normal load for the particle-substrate contact. 
}
\label{fig:a}
\end{figure}

\new{
The total normal forces $f^{\n}$ can be decomposed into (negative) adhesive forces, $-\h^{\n}$, and forces due to reversible contact deformation, $f_{\el}^{\n}(\delta^{\n})$, which are usually repulsive (i.e. positive), and velocity-dependent forces, $f_{\diss}^{\n}(\delta^{\n},v^{\n})$, that can take any sign, but vanish in static cases, so that}
\begin{subequations}\label{eq:normal}
\begin{gather}
f^{\n} = -\h^{\n} + f_{\el}^{\n}+f_{\diss}^{\n}. 
\end{gather}
\new{Note that the minus sign is written explicitly in front of the adhesive force, thus $\h^{\n}$ is positive.}

This model can be extended to model more complex effects such as elasto-plastic deformation\,\cite{Luding2008c,Tomas20075925,tykhoniuk2007ultrafine}, or various other interaction mechanisms -- however, we are restricted ourselves to elastic Hertz contacts since they are valid in most of the experiments for the materials, sizes and compression forces described in chapter \ref{sec:expr}. 
%
Elastic deformations of spheres are described by a Hertzian contact normal force\,\cite{hertz1882b}%
\[
f_{\el}^{\n}(\delta^{\n}) = \frac{4}{3} E^* \sqrt{\r} (\delta^{\n})^{3/2}, 
\label{eq:Hertz}
\]
where the modulus $E^*=\left[\frac{(1-\nu^2)}{E}+\frac{(1-\nu_\alpha^2)}{E_\alpha}\right]^{-1}$ is a combination of the  Young's moduli and Poisson's ratios of the particle ($E$, $\nu$) and the contacting surface ($E_\alpha$, $\nu_\alpha$), respectively, where \new{$\alpha=\i$} denotes the indenter and \new{$\alpha=\s$} the substrate, and the contact radius, see Figure \ref{fig:a}, is
\[a \approx\sqrt{R\delta^{\n}}\mbox{ for }\delta^{\n}\ll R.\label{eq:a}\] 
The adhesive force, $\h^{\n}$, 
can be measured as the pull-off force required to detach a particle from a surface. Since the contacts do not open, we do not specify non-contact adhesive force contributions here.

The dissipative force acts against the relative velocity and models the energy loss during a normal collision. 
\[f_{\diss}^{\n}= - \gamma^{\n} v^{\n},\]
\end{subequations}
with viscous dissipation factor $\gamma^\n(\delta^\n)$, which can be a function of both normal velocity and displacement \cite{brilliantov1998rolling,luding1998collisions,kuwabara1987restitution}.
For the simulations here, the dissipation coefficient was chosen according to \cite{thornton2012investigation} as $\gamma^{\n}=0.0062\sqrt{m k^{\n}}$, \read{with spring stiffness $k^\n=\frac{\partial f_\el^\n}{\partial\delta^\n} = 2E^* a$,} resulting in a constant restitution coefficient of $e=0.97$ \cite{Luding1998}. 
For simplicity, \new{and because we consider small velocities and thus vanishing dissipative forces,} the dissipation coefficients used in tangential direction are equal to the normal dissipation coefficient, but could differ in general.

\subsubsection{Tangential forces and torques}

\new{The tangential sliding force and the rolling and torsion torques are assumed to resist the relative tangential and angular velocities, $\vec{v}^\nu$, and act against displacements, $\vec{\delta}^\nu$, with $\nu=\l,\ro,\to$.
They can be} modeled using an elastic force coupled with a dissipative force,%
\begin{subequations}
\begin{gather}
\vec{f}^{\nu} = \vec{f}_{\el}^{\nu}(\vec{\delta}^\nu) + \vec{f}_{\diss}^{\nu}(\vec{v}^\nu),~\nu=\l,\ro,\to. \label{eq:Lineart}
\end{gather}

\new{%
The rolling and torsion \lq\lq{}forces\rq\rq{} calculated here are not actual forces, but only used as intermediates to calculate the rolling and torsion torques,
\begin{gather}
\vec{M}^{\ro} = \frac{\la^{\ro}}{r} \vec{r} \times \vec{f}^{\ro},\quad
\vec{M}^{\to} = \la^{\to} \vec{f}^{\to},
\end{gather}
}%
\end{subequations}
\new{%
where $\la^{\ro}$, $\la^{\to}$ have to be specified.

The dissipative forces act against the relative velocities $\vec{f}^{\nu,\diss}= - \gamma^\nu \vec{v}^\nu$. 
The elastic forces are determined by a spring stiffness, $k^\nu$ such that
\[ \vec{f}_{\el}^{\nu} = k^{\nu}\ \vec{\delta}^{\nu}.\]
If the spring stiffness varies during tangential displacement, the tangential force has to be computed incrementally \cite{thornton2012investigation}.
%
For Hertzian contacts, the Mindlin\,\cite{mindlin1949compliance} model is assumed a non-linear sliding force with a spring stiffness depending on the contact radius (and thus on the normal load),
\[ k^{\l,\textrm{M}}=8G^*a,\]
where $G^*=\left[\frac{(2-\nu^2)}{G}+\frac{(2-\nu_\alpha^2)}{G_\alpha}\right]^{-1}$ is a combination of the shear moduli $G$, $G_\alpha$ and Poisson's ratios $\nu$, $\nu_\alpha$ of the two materials. %
}%
This contact model is a simplification of the contact model of  \MD\,\cite{mindlin1953elastic}, implemented in\,\cite{thornton2012investigation}, where a varying spring stiffness $k^{\l,\mathrm{MD}}$ is used that depends on the normal and tangential displacement and the contact history (loading, unloading, reloading), resulting in a decreasing stiffness before yield. 
\new{Assumptions made for $k^\ro$, $k^\to$ will be discussed in section \ref{sec:rt}.}

\subsubsection{Yield criterion}
Each of the three tangential forces/torques is assumed to have an \emph{independent} yield criterion for slip:
When the ratio of tangential to normal force exceeds the static contact friction coefficient, $\mu^{\l,\s}$, the contact surface yields until the ratio becomes smaller than or equal to the dynamic contact friction coefficient, $\mu^{\l}$. 
This is modeled by a yield criterion, truncating the magnitude of $\vec{\delta}^{\l}$ as necessary to satisfy 
\begin{subequations}\label{eq:yield}
\[|\vec{f}^{\l}| \le \mu^{\l} (f^{\n}+\h^{\l})\label{eq:yield1}.\] 
The sliding friction coefficient $\mu^{\l}$ is usually \new{assumed} to be constant (Coulomb type).
%
For zero normal load, the adhesive force, $\h^{\l}$, equals the pull-off force $\h^{\n}$. 
%
%
While the static friction coefficient \new{is generally larger} than the dynamic friction coefficient, we assume for simplicity that $\mu^{\l,\s}=\mu^{\l}$.

Similarly, when the rolling- (or torsion-) torque-to-normal-contact-force ratio becomes larger than the contact friction coefficient, $\mu^{\nu}$, the magnitude of $\vec{\delta}^{\nu}$ is cut as necessary to satisfy 
\[|\vec{M}^{\nu}| \le \mu^{\nu} \la^{\nu} ({f}^{\n}+\h^{\nu}),\ \nu=\ro,\to\label{eq:Myield}.\]
\end{subequations}
where the length scale, $\la^{\nu}$, will be discussed and specified later.
\new{Note that the friction coefficients $\mu^{\nu}$ and adhesive forces $\h^{\nu}$ are not necessarily constant, but can depend, \emph{e.g.}, on the particle radius, the contact radius, or speed.}
But even though they are variable quantities, we set both to be constant for the sake of a simple contact model. Thus, $\h^{\nu}$ did not necessarily represent the actual adhesive force, but simply defines the offset for the yield criteria, see Eqs.\,\eqref{eq:yield}. 
\new{Alternatively to the separate yield criteria, combined failure criteria can be derived from the assumption of a perfectly cylindrical solid bridge, see \eg \cite{BrendelTorokKirschBrockel2011}.}

\subsubsection{Relation to literature} 
\label{sec:rt}

Many effects such as surface roughness, adhesion, and liquid/solid bridges can cause rolling and torsion \quotes{friction}, and thus influence the stiffness $k^{\nu}$, the friction coefficients $\mu^{\nu}$ and the length scale $\la^{\nu}$, $\nu=\ro,\to$. 
\new{Note that the split of the friction factors into coefficients and lengths is for discussion/comparison of and with other models, while in the simulation models, the friction factor is used.}
For example, in\,\cite{yang2000computer} the rolling friction coefficient was assumed to be constant when scaled by the particle radius, $\la^{\ro}=r$, based on elastic hysteresis losses\,\cite{beer2003vector}. 
%
In\,\cite{BrilliantovPoschel1998} a velocity-dependent rolling friction coefficient $\mu^{\ro}\propto v^{\ro}$ with length scale $\la^{\ro}=R$ was derived from viscoelastic deformations, which was independent of the contact area.
In\,\cite{Tomas20075925}, rolling for a Hertzian normal contact was assumed to be caused by micro-creep\,\cite{johnson1984contact}, yielding a constant friction coefficient scaled by the contact radius, 
\begin{subequations}
\[\mu^{\ro,\mathrm{Tomas}}=1\mbox{ for } \la^{\ro}=a. \label{eq:muro_to}\]
In the same paper, torsion is calculated based on the Deresiewicz model\,\cite{deresiewicz1954contact}, yielding a torsional friction coefficient proportional to the sliding friction coefficient, 
\[\mu^{\to,\mathrm{Tomas}}=(2/3)\mu^{\l}\mbox{ for } \la^{\to}=a. \label{eq:muto_to}\]
\end{subequations}
Thus, both friction factors \new{can depend on each other via} the contact area.
A similar prediction for the torsional friction coefficient, $\mu^{\to,\mathrm{Dintwa}}=(3\pi/16)\mu^{\l}\mbox{ for } \la^{\to}=a$ has been made in\,\cite{dintwa2005torsion}, using a different yield condition.
\new{Furthermore, Farkas et al. \cite{Farkas2003}, who  studied the interplay between torsion and sliding friction, found that the presence of torsion can reduce the sliding friction; this effect, however, is not considered here.}

\subsection{DEM implementation of rolling, sliding and torsion experiments}

The sliding, rolling and the rail (combined rolling and torsion) experiments will now be analyzed and compared to simple \new{single-particle} DEM simulations.
For a sketch of the forces and torques acting on the particles in the simulation, see \figref{fig:forces0}. In all three experiments, a coordinate system is used where $\e{x}$ denotes the shear direction, $\e{y}$ the vertical axis and $\e{z}=\e{x}\times\e{y}$ the cross-shear direction. 
In each simulation, a constant normal load $\vec{f}_{\i}^{\n}$ is applied to the indenter and the system is equilibrated until the normal forces are in balance. Then the indenter is sheared back and forth between $x=\pm1\,\mathrm{\mu{}m} $ at constant speed $|\vec{v}_{\i}|=1\,\mathrm{\mu{}m/s}$ and the shear force $f_{\i}^{\l}$ acting at the indenter in $x$-direction is measured. 
For simplicity, we consider the same elastic stiffness for the rolling, torsion and sliding torques, $k^{\ro}=k^{\to}=8G^*a$.

\section{Experimental results and interpretation} \label{sec:expr}

In this section, the distinct differences of the effects of roughness and adhesion on sliding and rolling of micron-sized borosilicate particles on Si surfaces are explored. \new{After the roughnesses are determined in section \ref{sec:41}, normal, sliding and rolling tests are performed and analysed in sections \ref{sec:43}-\ref{sec:46}. Furthermore, by use of rail structured Si surfaces, a first experimental distinction between rolling, sliding and torsion is reported in sections \ref{sec:47}-\ref{sec:48}.} All experimental results are compared with analytical results and DEM simulations. 

\begin{table*}[tbp]
\centering
\caption{Material properties for substrates, particle and probes used for normal, sliding, rolling and torsion tests
\label{tab:mat}
}
\begin{tabular}{lcccc}
\hline\hline
 & Substrate 1 & Substrate 2 & Indenter & Particle\\\hline
Material & Si(100)& fused quartz & Diamond & Borosilicate \\
Young's modulus & $E_{\s}=179\unit{GPa}$ & $E_{\s}=71\unit{GPa}$ & $E_{\i}=1140\unit{GPa}$ & $E=71\unit{GPa}$\\
Poisson's ratio & $\nu_{\s}=0.25$ & $\nu_{\s}=0.17$& $\nu_{\i}=0.07$ & $\nu=0.17$ \\
RMS roughness\,$\Delta$\,[nm]$\!\!\!\!\!\!$ & see Tab.\ \ref{tab:surf} & & $4.0\pm1.0$ & $0.7\pm0.1$\\
 & (for sliding/rolling/rail) & (for normal testing) & (for rolling/rail) & $R=8.5\unit{\mu{}m}$\\ 
\hline
\end{tabular}
\end{table*}

\begin{table*}[tbp]
\centering
\caption{Contact properties for sliding and rolling tests on different rough Si surfaces and effective friction values for rolling/torsion tests. 
\label{tab:surf}
}
\renewcommand{\tabcolsep}{3pt}
\begin{tabular}{lcccl}
\hline\hline
Substrate 1 surfaces & untreated & etched (1600\,W) & etched (1800\,W) & section\\\hline
RMS roughness $\Delta$\,[nm] & $0.3\pm0.1$ & $1.5\pm0.2$ & $2.7\pm0.4$ & \multirow{2}{*}{\bigg\} section \ref{sec:41}}\\
Peak-to-valley\,[nm] & $1.0\pm0.1$ & $10.5\pm0.7$ & $14.8\pm1.0$ \\
AFM\,pull-off\,force\,$\h^{\n}$\,[$\unit{\mu{}N}$]\!\!\!\!\!
& $3.2\pm0.3$ & $2.7\pm0.3$ & $1.9\pm0.3$ & \multirow{3}{*}{\Bigg\} section \ref{sec:43}}\\
Sliding friction $\mu_{\s}^{\l}$ & $0.23\pm0.005$ & $0.53\pm0.005$ & $0.65\pm0.014$\\
Sliding adhesion $\h_\s^{\l}$\,[$\unit{\mu{}N}$] & $12.00\pm1.20$ & $4.45\pm0.55$ & $7.76\pm1.17$\\
Rolling fr. $\bar{\mu}^{\ro}$\,for\,$\bar\la^{\ro}\!=\!\bar r$ & ($7.9\pm0.3\!)\cdot\!10^{-4}$ & ($4.6\pm0.2\!)\cdot\!10^{-4}$ & ($3.5\pm0.3\!)\cdot\!10^{-4}$ &  \multirow{4}{*}{$\left.\rule{0cm}{.8cm}\right\}$ section \ref{sec:45}}\\
Ro.\,adh.\,$\bar\h^{\ro}$\,for\,$\bar\la^{\ro}\!=\!\bar r$\,[$\unit{\mu{}N}$]  & $571.4\pm56.1$ & $1398.0\pm82.7$ & $2029.4\pm149.2$\\
Rolling fr. $\bar{\mu}^{\ro}$\,for\,$\bar\la^{\ro}\!=\!\bar a$ & ($7.6\pm1.0\!)\cdot\!10^{-3}$ & ($1.8\pm0.7\!)\cdot\!10^{-3}$ & ($1.7\pm1.3\!)\cdot\!10^{-3}$\\
Ro.\,adh.\,$\bar\h^{\ro}$\,for\,$\bar\la^{\ro}\!=\!\bar a$\,[mN]  & $2.6\pm0.6$ & $13.5\pm5.6$ & $12.6\pm11.0$\\
\\\hline\hline
Rail inclination $[{}^\circ]$ & $25\pm0.5$ & $45\pm0.5$ & $65\pm0.5$ \\\hline
RMS roughness $\Delta$ [nm] & $2.5\pm1.0$ & $3.6\pm1.4$ & $4.8\pm2.0$ & \phantom{\}\} }section \ref{sec:41}\\
Rail friction coeff. $\mu^{\rail}$ & ($1.58\pm0.02)\cdot10^{-3}$ & ($2.42\pm0.09)\cdot10^{-3}$ & ($3.34\pm0.19)\cdot10^{-3}$ & \multirow{4}{*}{$\left.\rule{0cm}{.8cm}\right\}$ section \ref{sec:47}}\\
Rail adhesion $\h^{\rail}$\,[$\unit{\mu{}N}$] & $173.4$ & $57.8$ & $191.3$\\
Torsion fr. $\mu_{\i}^{\to}$\,for\,$\la_\s^{\to}\!=r_\s\!$  & & ($5.6\pm0.5)\cdot10^{-3}$ \\
Torsion fr. $\mu_{\i}^{\to}$\,for\,$\la_\s^{\to}\!=a_\s\!$ & & ($6.4\pm1.2)\cdot10^{-2}$ 
\end{tabular}
\end{table*}

\subsection{Test objects}\label{sec:41}

\subsubsection{Silicon surface roughness}

\begin{figure*}[!ht]
\centering
\includegraphics[width=.8\textwidth]{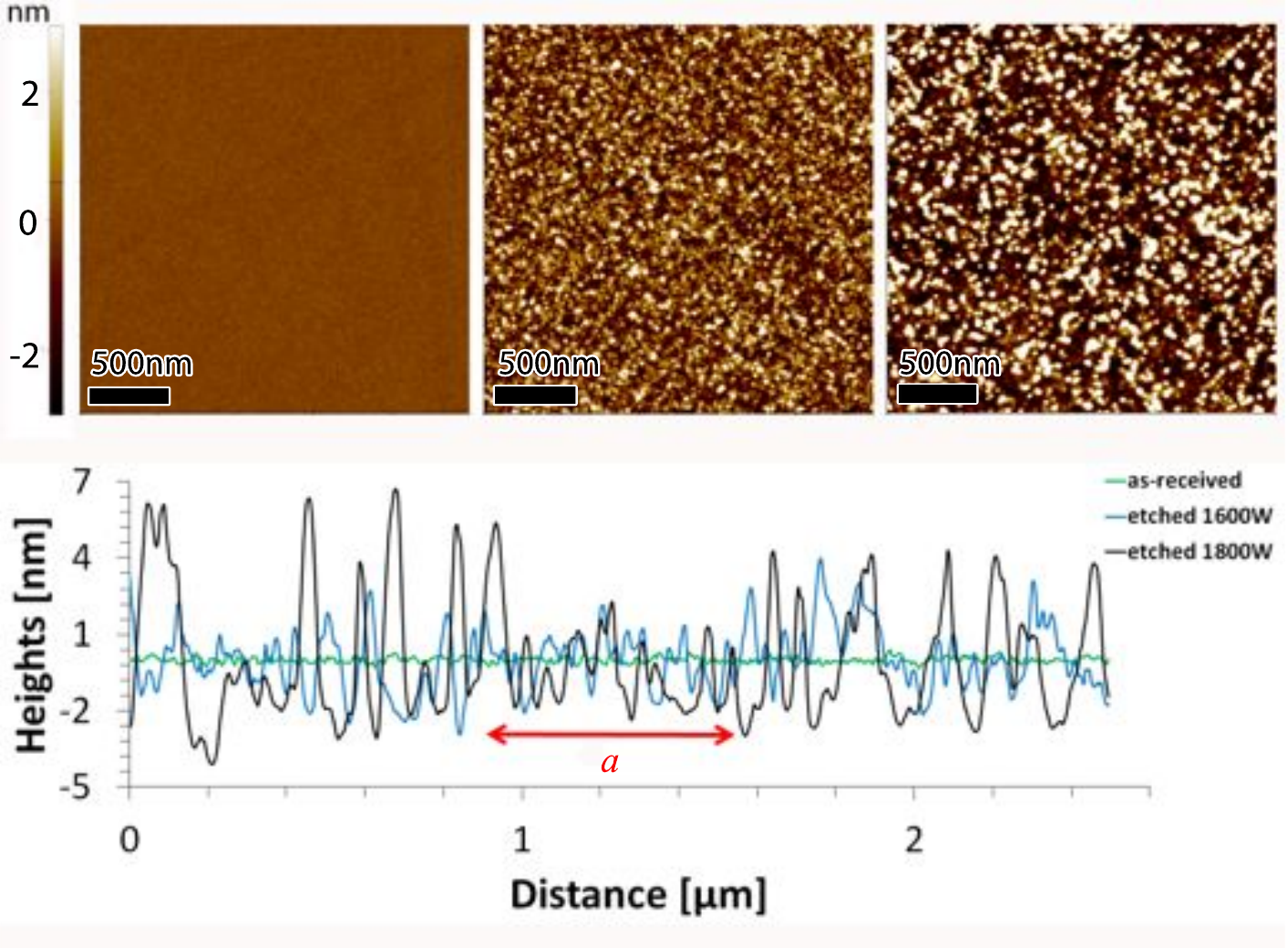}
\caption{AFM topography images (upper part) as well as cross section profiles (lower part) of the Si surfaces clearly show the roughness change after MWCVD H\textsubscript{2} etching with 1600\,W (middle) and 1800\,W (right) power. The RMS variations of the surface heights are $0.3\pm0.1\,$nm, $1.5\pm0.2\,$nm and $2.7\pm0.4\,$nm (upper panels, from left to right). 
For comparison, a length of $0.50\unit{\mu{}m}$ is shown, which corresponds to the radius of a particle-substrate contact calculated with Hertz theory at a normal load of $1000\unit{\mu{}N}$ and is larger than the typical horizontal peak-to-valley distances, but not much larger than the asperities, see Table~\ref{tab:mat}. }
	\label{fig:AFM_MWCVD_Si}
\end{figure*}

In \figref{fig:AFM_MWCVD_Si}, upper panels, one example AFM topography image (scanned area $2.5\times2.5\,\mathrm{\mu m^{2}}$) of differently treated and untreated surfaces are given. A clear change in topography of the Si samples with respect to increasing microwave power is visible from left to right. This becomes more obvious in the cross sectional analysis of the asperity heights versus the scan line, shown in the lower panel.
Correspondingly, RMS roughnesses of $0.3\pm0.1\,$nm, $1.5\pm0.2\,$nm and $2.7\pm0.4\,$nm are measured for microwave powers of 0\,W, 1600\,W and 1800\,W (\tabref{tab:mat}), respectively . It turns out that the plasma etching process is capable of systematically changing the RMS roughness of the Si surfaces. As expected, a higher microwave power leads to an increase in surface damage and in turn to a higher final roughness values. In this context, the distribution of surface heights features Gaussian characteristics for all three sample surfaces (not shown), indicating the random nature of this particular etching process.

\new{The large, but not too large, ratio of contact radius (Hertz theory) and RMS roughness} of the borosilicate sphere (double-sided red arrow, lower panel) at a normal load of 1\,mN indicates that \new{the effect of different surface roughnesses} cannot be ignored. 

\subsubsection{\read{Silicon adhesion}}\label{sec:adh}

The results of the colloid probe technique based adhesion measurements are summarized in \tabref{tab:surf}.
The measured adhesion forces, $\h^{\n}$, decrease with increasing RMS roughness. 
As surface chemistry (amorphous SiO\textsubscript{2}) and measurements conditions are kept constant, this finding is directly related to a diminished real contact area due to the roughness. The results are in good agreement with recent experimental results of Liu et al.\,\cite{Liu2007}, who studied the correlation between adhesion, tip radius and surface roughness in more detail.

\subsubsection{Rail system roughness and angle}

The surface roughness of the rails,  see \tabref{tab:surf}, varies with rail  inclination. Larger  inclinations show increased values due to the preparation procedure by FIB. 
The  inclination  was characterized by FESEM based cross-sectional analysis, which was able to confirm the target inclination of $25\degree$, $45\degree$ and $60\degree$, respectively, \new{with an error of $0.5^\circ$}.

\subsection{Normal testing of nanoindentation based colloid probes}\label{sec:42}

\begin{figure}[bp]
	\centering
		\includegraphics[width=1\columnwidth]{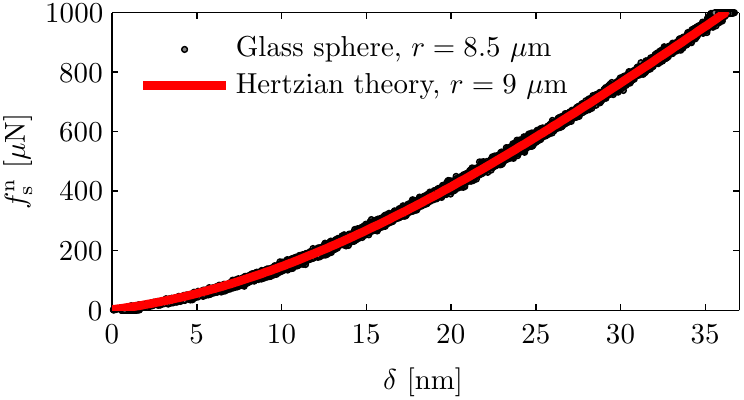}
		\caption{Normal load $f_{\s}^{\n}$ vs. displacement curve $\delta$. The red line shows a fit with Hertzian theory by assuming a radius of $9\,\mathrm{\mu m}$, with materials from \tabref{tab:mat}.}	
	\label{fig:normal_1}
\end{figure}

\figref{fig:normal_1} shows a load-displacement curve obtained by using a nanoindenter colloid probe with a radius of \read{$8.5\,\mathrm{\mu m}$} on a silica substrate. Such measurements were carried out in the context of the probe evaluation after glueing and radius determination with the FESEM. Elastic modulus and Poisson's ratio are assumed to be 71\,GPa and 0.17 for both sphere and surface (\tabref{tab:mat}), respectively. A simple Hertzian fit with a slightly larger radius of $9\,\mathrm{\mu m}$ was able to \new{accurately} reproduce the experimental data. The slightly stiffer response has to be attributed to a combination of surface roughness (about $0.7\pm0.1\unit{nm}$) and non-sphericity (i.e. local variation in actual radius). Thus, a correction to the modulus of the sphere by a factor of 1.05 (keeping a constant sphere radius of $8.5\,\mathrm{\mu m}$) can also be chosen here, which has the same effect like a slightly larger radius of $9\,\mathrm{\mu m}$. 

According to Tabor\,\cite{Tabor1996}, the yield stress can be approximated as one third of the hardness. For the silica surface used in this experiment, a measured hardness of 9\,GPa results in a yield stress of 3\,GPa, slightly \read{above the maximum contact pressure of $2.8$\,GPa acting at the contact area with a flat substrate, 
based on the highest normal loads used in our sliding and rolling experiments, $f_\s^\n\leq3000\,\mu\textrm{N}$.} Therefore, with silica being the weakest material in this study, we have a predominantly elastic Hertzian contact in all experimental settings presented here. 
\read{However, the pressure at individual small surface asperities will be sufficient to initiate plastic deformation\read{\,\cite{Bhushan1998}}. Thus, surface roughness and plastic deformation cannot be ignored for the other degrees of freedom, as shown below.}

\subsection{Sliding tests}\label{sec:43}

\begin{figure}[bp]
	\centering
		\includegraphics[width=1\columnwidth]{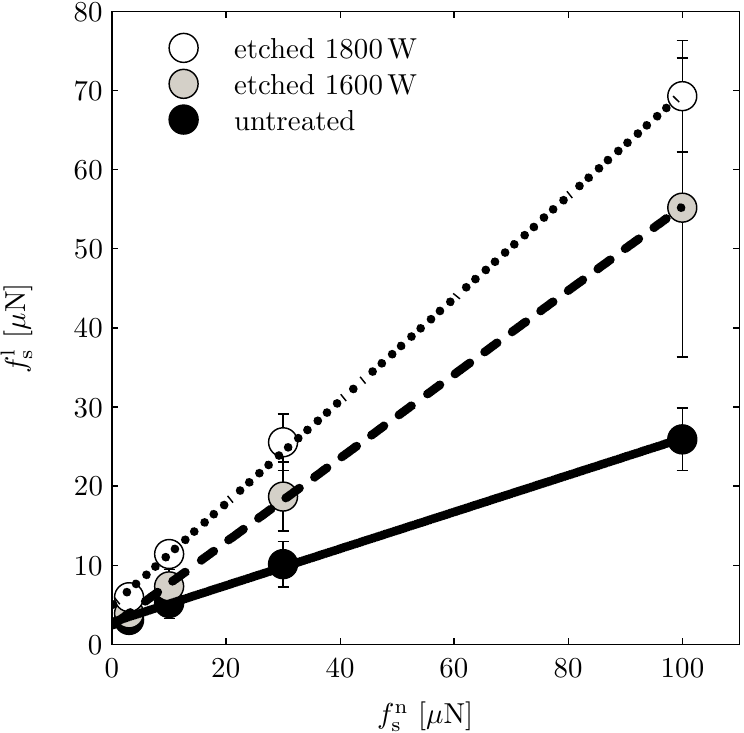}
		\caption{Resulting lateral force $f_{\s}^{\l}$ vs.  normal load $f^{\n}_{\s}$ for sliding friction measurements. The lines represent simple linear fits to the corresponding data sets, from which $\h_\s^\l$, $\mu_\s^\l$ are extracted.}
	\label{fig:sliding_1}
\end{figure}

The results of the sliding friction tests are summarized in \figref{fig:sliding_1}. A linear dependence between the measured lateral force $f_{\s}^{\l}$ and the applied normal load $f_{\s}^{\n}$ can be seen. The linearity of these three curves indicates that the modified Coulomb friction law \eqref{eq:yield1} holds for the given experimental conditions. A linear fit of the data yields the sliding friction coefficient $\mu_\s^{\l}$ along with an offset, which can be related to the adhesion  $\h^{\n}$ between the contact partners. The results from the linear regression are summarized in  \tabref{tab:surf}.

None of the tests, which have been carried out within the framework of the present study, showed any time dependent behavior that potentially could be attributed to the \new{much softer, viscous} layer of glue between particle and diamond tip. 
However, in case of imperfect gluing we see effects for significantly higher applied normal loads (5-10\!mN).

The second observation is in \new{qualitative} agreement with the decrease in adhesion measured by the AFM pull-off force, $\h^\n$, for the contact partners. 
\read{The larger values of sliding adhesion forces gained from the tribological characterization as compared to the  adhesion measurements in section \ref{sec:adh}}, however, potentially stem from the differences in dynamics of the process that detaches micro-asperities from a surface and the process that initiates \read{new contacts during} shear of the sphere along the substrate.

The first observation, however, is somewhat surprising as it hints towards a substantial onset of irreversible plastic deformation in the contact zone even at the relatively low normal loads applied during the tests. A simple Hertzian contact with perfectly flat surfaces, even though predicting the normal force well, is not able to explain any onset of plasticity. Nevertheless, the pressure at individual small surface asperities will certainly be sufficient to initiate plastic deformation\read{\,\cite{Bhushan1998}}. 

The experimental clarification and/or verification of this particular issue, i.e. the determination of changes in surface topography either on the sphere or the surface area in contact, is work in progress and will be reported elsewhere.

\subsection{Comparison to sliding simulations}\label{sec:44}

\begin{figure}[b]
\centering
\includegraphics[width=1\columnwidth]{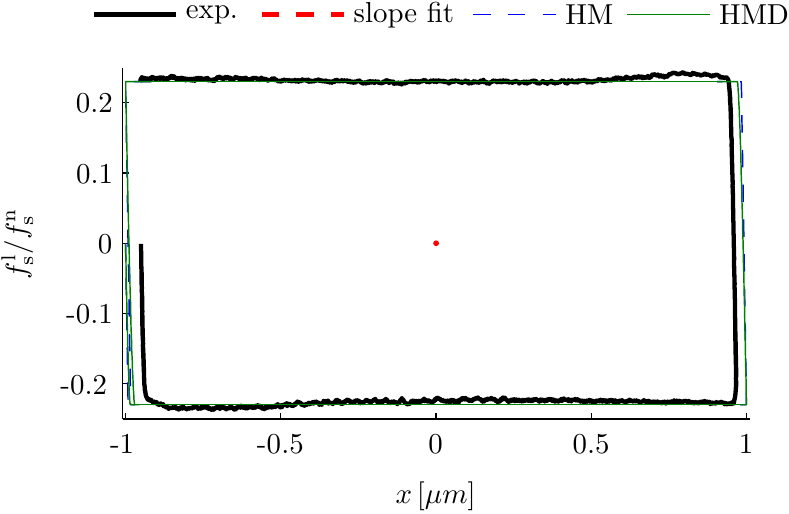}
\setlength{\unitlength}{\columnwidth}
\begin{picture}(0,0)
\put(-.18,.22){\includegraphics[width=.48\columnwidth]{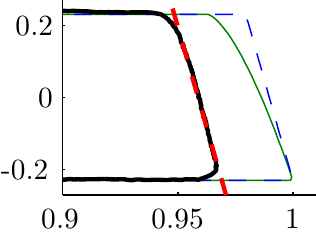}}
\end{picture}
\caption{Shear force plotted against displacement for the sliding experiment shown in \figref{fig:forces0}a, for the untreated substrate and $f_{\s}^{\n}=100\unit{\mu{}N}$ (thick black line). This is compared to DEM simulations using the Hertz-Mindlin model (HM, blue dashed line) and Hertz-\MD{} (HMD, green line), with a fitted shear modulus of $G^{*,\mathrm{fit}}=1.21\unit{GPa}$ (thick red dashed line).
The difference between experimental and numerical reversal points is on purpose to improve visibility.
}
\label{fig:slide}
\end{figure}

For the sliding case, shown in \figref{fig:forces0}(a), the particle is attached to the indenter and thus moved with the same velocity as the indenter.
The resulting shear force vs. displacement curve is compared to an experimental friction loop in \figref{fig:slide}. 
\read{The friction loop shows} nearly linear elastic behavior for short intervals after the reversal point (where the velocity switches directions). For the elastic behavior, the force balance yields that the slope of the shear force vs. displacement curve at the load reversal point equals the negative sliding spring stiffness, ${d\,f_{\el,\i}^{\l}}/{dx}=-k^{\l}$. For both the linear Mindlin and the nonlinear \MD\, \cite{mindlin1953elastic} models, a stiffness of $k^{\l}=8G^* a$ holds true at the reversal point.

\read{The slope fitted to the experimental friction loop plotted in \figref{fig:slide} yields $k^{\l,\mathrm{fit}}=2.219\unit{kN/m}$ for a normal load of $f_{\s}^{\n}=100\unit{\mu{}N}$; this yields $G^{*,\mathrm{fit}}=k^{\l,\mathrm{fit}}/(8a)=1.36\unit{GPa}$}, which is much smaller than the value calculated from the material parameters in \tabref{tab:mat} of a perfectly flat borosilicate sphere and a silicon substrate ($G^*=11.87\unit{GPa}$). This result can be due to the nanostructure of the surfaces; however, it is difficult to interpret as the elasticity of the nanoindenter setup could possibly influence the slope observed in the friction loop. 

After the sliding spring yields, the magnitude of the sliding force remains constant at $\mu_{\s}^{\l} |f_{\s}^{\n}|$. Fitting the experimental results for the untreated surfaces shows a constant friction coefficient $\mu_{\s}^{\l}\approx0.23$. 
\read{A linear fit of the lateral force against the contact area, $f_\s^\l\propto \pi a^2 \propto (f^\n + \h^\n)^{2/3}$, as suggested in \cite{tykhoniuk2007ultrafine}, did not produce good, consistent results.}
For higher roughness, and a peak at the transition from the elastic to the yielding sliding force (not shown) suggests that the static sliding friction coefficient is higher than the dynamic one. However, this will not be discussed further here.

\subsection{Rolling tests}\label{sec:45}

\begin{figure}[bp]
	\centering
		\includegraphics[width=1\columnwidth]{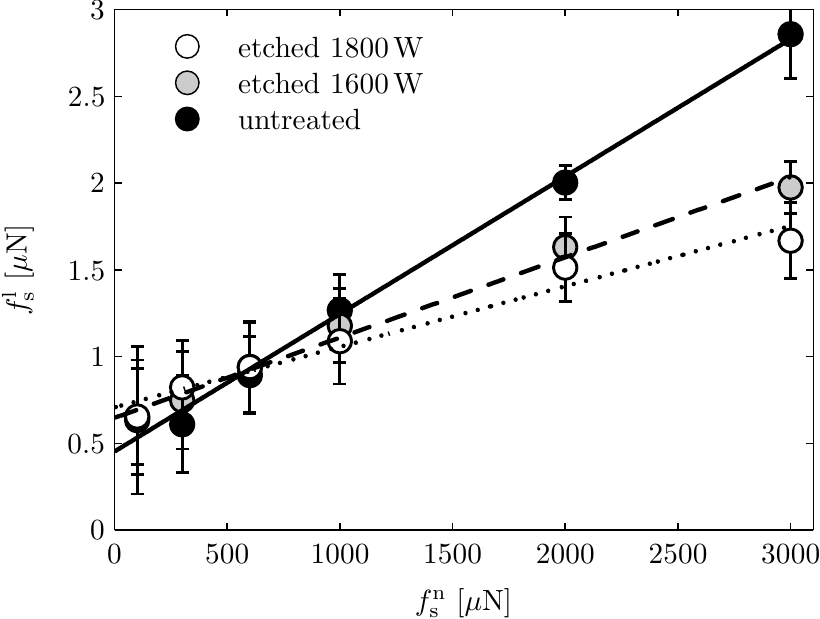}
		\caption{Resulting lateral force  $f_{\s}^{\l}$ vs. normal load $f^{\n}_{\s}$ for rolling friction measurements. The lines represent simple linear fits to the corresponding data sets.
}
	\label{fig:rolling_1}
\end{figure}

\figref{fig:rolling_1}  shows the results of the rolling experiments. Again, a straight line fits the data although some additional scattering is due to small variations of the radii of the different spheres used. Processing the experimental data in an analogous manner to the sliding tests yields the \new{mean} rolling friction coefficient $\bar\mu^{\ro}$, \new{as described in detail in section \ref{sec:46}}. The measured lateral force signals are 2-3 orders of magnitude smaller for rolling compared to sliding tests, which confirms that the particle is rolling over the surface. Surprisingly, in contrast to the sliding data, $\bar\mu^{\ro}$ decreases with increasing surface roughness, while the intercept with the abscissa shifts to the left\read{, \ie rolling adhesion $\bar\h^\ro\gg\h_\s^\l$ increases with surface roughness}. The latter finding suggests that a simple interpretation of this intercept as a measure of the normal adhesion force between the contacting partners is not straightforward in the case of a rolling contact.

Ideally, a non-adhering rolling sphere contacts with the underlying surface at only one point, which would result in a zero rolling friction 
\new{since no torque can be  created by a point contact.}
%

In our system, the finite contact surface, involving many single asperities on the sphere as well as on the substrate surface, leads to a non-zero rolling resistance due to various (three at least) dissipative phenomena.
Rolling is a continuous generation of new contact area at the front and a continuous contact detachment at the rear of the contact (viewed in rolling-direction of the sphere).
During this process, multiple mechanisms can dissipate energy due to the rough nature of the surfaces \cite{Lothar}:
($i$) the formation and breaking of contacts between the asperities causes small oscillations/waves at the front and rear of the contact, which are viscously damped, proportional to the rolling speed; 
($ii$) in the presence of humidity, water (layers and) bridges lead to viscous effects, too; 
furthermore, liquid bridges form between the asperities at the front when touching, but break at the backside of the sphere at a larger separation, thus leading to a net rolling resistance by themselves, with additional effects due to a plastically changed surface;
($iii$) asperities deform plastically under normal load when new surface is created leading to a plastic rolling resistance; and finally, plasticity also causes asperities to be flatter at the rear than at the front, creating an asymmetric distribution of van der Waals forces, i.e. stronger attraction at the back than at the front.
For each of these mechanisms, the dissipated energy and thus rolling resistance is related to the surface structure (e.g. to the density and roughness amplitude of the asperities) of the Si surfaces. Therefore, rolling friction appears to be attributable to various combinations of viscous, wet, and plastic effects at the individual asperities, as supported by reports of other groups. One of the first overviews attributing rolling friction to adhesion was given by Briggs and Briscoe in 1976\,\cite{Briggs1976}; a more recent study has been presented by Sumer and Sitti\,\cite{Sumer2008}.

\subsection{Analysis of the rolling measurements}\label{sec:46}

For the rolling case shown in \figref{fig:forces0}(b), the particle is \new{not attached to} the indenter. Thus, it can rotate, and rolling torques are created.
The forces and the torques acting on the particle are
\begin{subequations}\label{eq:balance_ro}
\begin{gather} 
\vec{f} = \vec{f}^{\l}_{\i} + \vec{f}^{\l}_{\s} + \vec{f}^{\n}_{\i} + \vec{f}^{\n}_{\s}, \label{eq:forces1} \\
\vec{M} = \vec{r}_{\i}\times\vec{f}^{\l}_{\i} + \vec{r}_{\s} \times\vec{f}^{\l}_{\s} + \vec{M}^{\ro}_{\i} + \vec{M}^{\ro}_{\s}. \label{eq:torques1a} 
\end{gather}
\end{subequations}

%
%
To calculate the maximum sliding force, we assume that the system is in steady state (i.e. $\vec{f}=\vec{M}=\vec{0}$) to obtain from \eqref{eq:balance_ro}
\[ 2\vec{r}_{\s}\times \vec{f}^{\l}_{\s} = -\vec{M}^{\ro}_{\i} -\vec{M}^{\ro}_{\s}, \label{eq:torques1b} \]
using $\vec{r}_{\s}\approx -\vec{r}_{\i}$, while $\vec{f}_{\i}^{\n}=-\vec{f}_{\s}^{\n}$ and $\vec{f}_{\i}^{\l}=-\vec{f}_{\i}^{\s}$ follow from the components in \eqref{eq:forces1}.  
\read{These equalities allow us to use the simplified notation, $\bar{r}=(r_\s+r_\i)/2$, $f^\n=|\vec{f}_{\i}^{\n}|$ and $f^\l=|\vec{f}_{\i}^{\l}|$.
 
If we further assume $\mu_\alpha^{\ro}  \la_\alpha^{\ro} \leq \mu_\alpha^{\l} r_\alpha$, $\alpha=\s,\i$, such that the rolling forces yield before the sliding forces, then we obtain from \eqref{eq:Myield} and \eqref{eq:torques1b} 
\[2 \bar{r} f^\l=\mu_{\s}^{\ro} \la_{\s}^{\ro}({f}^{\n}+\h^{\ro}_{\s})+\mu_{\i}^{\ro} \la_{\i}^{\ro}({f}^{\n}+\h^{\ro}_{\i}). \label{eq:fs1}\]

\begin{subequations}\label{eq:barto}
Thus, the measured lateral force satisfies
\[f^{\l}= \frac{\bar\mu^{\ro}\bar\la^{\ro}}{\bar{r}} ({f}^{\n} + \bar\h^{\ro}). \label{eq:fs1b}\]
where the measured friction coefficient,
\[\bar\mu^{\ro}=\frac{\mu_{\s}^{\ro}\la_\s^\ro+\mu_{\i}^{\ro}\la_\i^\ro}{2\bar\la^\ro},\label{eq:barmuro}\] 
is the mean friction factor divided by the mean rolling length scale,
\[\bar\la^\ro=\frac{\la_\s^\ro+\la_\i^\ro}{2},\label{eq:barlaro}\] 
and the averaged rolling adhesion force is
\[\bar\h^{\ro}=\mu_{\s}^{\ro}\frac{\la_{\s}^{\ro}\h^{\ro}_{\s}+\mu_{\i}^{\ro} \la_{\i}^{\ro} \h^{\ro}_{\i}}{\mu_{\s}^{\ro} \la_{\s}^{\ro}+ \mu_{\i}^{\ro} \la_{\i}^{\ro}}.\label{eq:barhro}\] 
\end{subequations}}%
\read{The measured coefficients can be found in \tabref{tab:surf} for $\bar\lambda^\ro{=}\bar{r}$ and $\bar\lambda^\ro{=}\bar{a}{=}(a_\s+a_\i)/2$, where the contact radii  been obtained using \eqref{eq:Hertz} and \eqref{eq:a} and the data in \tabref{tab:mat}}. 
In both cases, the rolling adhesion force, $\h^{\ro}$, is much higher than the pull-off force $\h^{\n}$, as the rolling friction appears to be large even for zero normal load. 
As discussed above, $\mu_{\s}^{\l}$ increases with higher surface roughness while $\bar\mu^{\ro}$ decreases. These results show that $\bar\mu^{\ro}$ is dominated by surface and interfacial forces such as \emph{e.g.} water films while $\mu_{\s}^{\l}$ is dominated by mechanical frictional stick slip effects at asperities. 

\begin{figure}[bp]
\vspace*{-.1in}
\centering
\includegraphics[width=1\columnwidth]{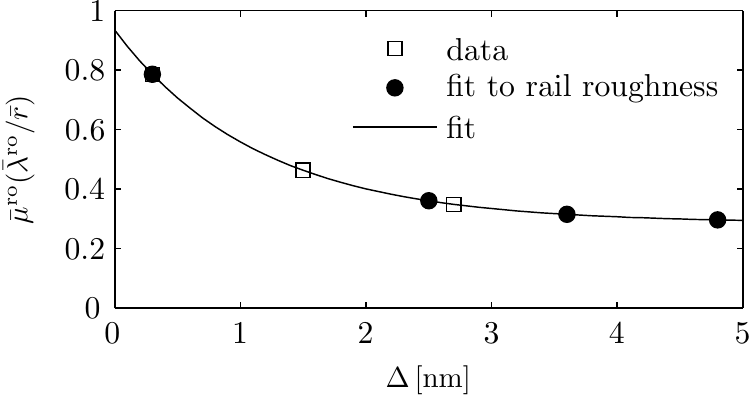}

\smallskip\includegraphics[width=1\columnwidth]{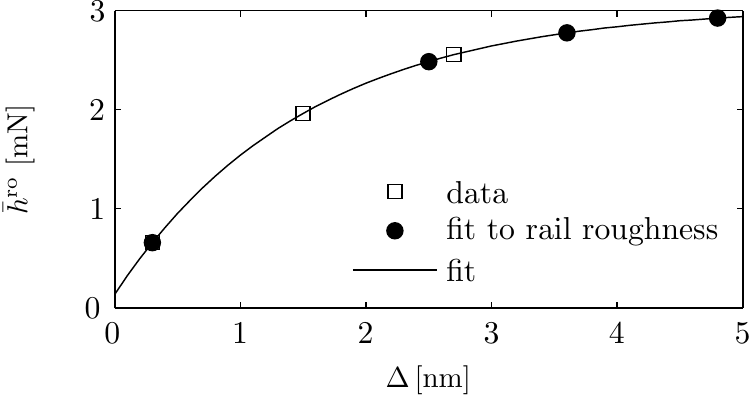}
\vspace*{-.1in}
\caption{Fit of the rolling friction coefficient $\bar\mu^{\ro}$ (top) and the rolling adhesion force $\bar\h^{\ro}$ (bottom) as functions of RMS roughness. This fit is used in section \ref{sec:48} to estimate the rolling friction in the rail experiments, based on the RMS roughnesses of the rail surface (dots).
}
\label{fig:rolling_2}
\end{figure}

To estimate the rolling friction in the following rail experiments, \read{both friction factor and rolling adhesion are} fitted against the RMS roughness of the substrate, $\Delta$, as an exponentially decaying value,
\begin{subequations}\label{eq:rofit}%
\begin{equation}
\bar{\mu}^{\ro}\frac{\bar\la^\ro}{\bar{r}} = b+c\exp(-\Delta/d),\label{eq:murofit}
\end{equation}
and
\begin{equation}
\bar\h^{\ro} = b'{-}c'\exp(-\Delta/d').\label{eq:frofit}
\end{equation}
\end{subequations}
with $b =   0.0002859$, $c =   0.0006475$, $d=1.1548\,$nm
and $b' =3.046\,$mN, $c' = 2.907\,$mN, $d'=1.5216\,$nm, as shown in \figref{fig:rolling_2}. The rolling friction factors in the rail experiments in section \ref{sec:48} are then assumed to satisfy the friction factors fitted to the rail roughness, see \tabref{tab:surf}.

The exponentially decaying value is in good agreement with the adhesion--roughness dependency\linebreak[4] shown by Liu et al.\,\cite{Liu2007}. This does not only suggest that the rolling seems to be adhesion dominated but also that a single-asperity model can be used to describe the contact behavior.

The measured rolling friction factor is much lower than the contact radius $a_\alpha$, in contrast to what was predicted in \eqref{eq:muro_to}. \new{However, as the data can be fitted well for both $\lambda^\ro=\bar r$ and $\bar a$,} 
a dependence of the friction factor on the normal force cannot be ruled out. This will be studied further in the future. 

Finally, no velocity dependence was observed for indenter velocities  $|\vec{v}_{\i}|=0.01,0.1,1\,\mathrm{\mu{}m/s}$, thus the rolling friction is not dominated by viscous effects \cite{BrilliantovPoschel1998} in this regime.

\subsection{Combined rolling and torsion rail tests}\label{sec:47}

From a mechanical point of view, torsion friction is due to a rotation of the sphere in normal direction and rolling friction is due to a rotation in tangential direction to the substrate. 
Experimentally, a controlled particle surface motion is possible by means of a Si based rail system featuring different  inclinations. 
\read{In the rail, these motions occur simultaneously which results in an effective friction coefficient with contributions from both torsion and rolling resistance; for details, cf. \eqref{eq:murail}.}

\read{Sliding motion at the indenter occurs for normal loads below a minimum normal load of $f_\i^\n=100\,\mathrm{\mu N}$, which is required to overcome the critical rolling resistance, based on the optical images which show no change in position after the rolling test for $f_\i^\n<100\,\mathrm{\mu N}$ (see section \ref{sec:24}). 
Above this threshold, pure rolling and torsion motion are active.
With higher inclination of the rail, the torsion contribution increases whereas the rolling part becomes comparatively smaller, as shown later in \eqref{eq:forces6}.}

\begin{figure}[t!]
	\centering
		\includegraphics[width=1\columnwidth]{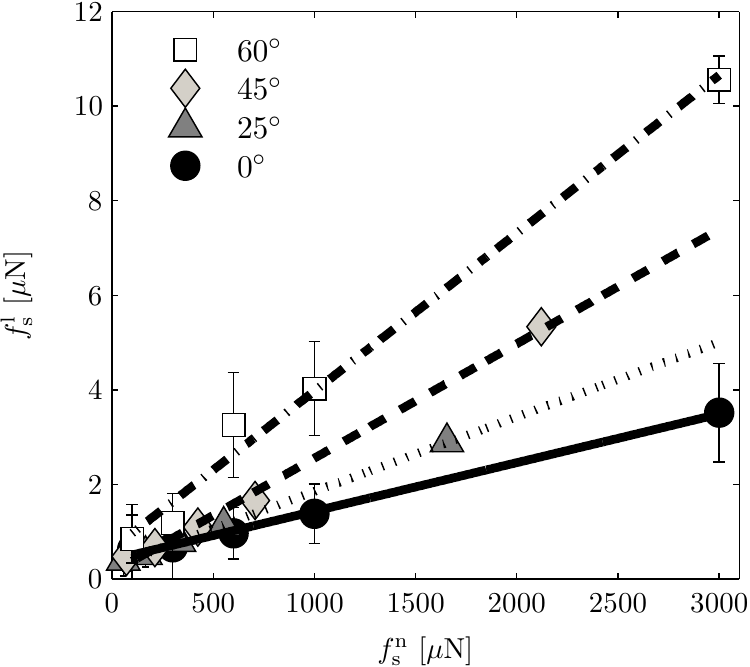}
		\caption{Resulting lateral force $f^{\l}_\s$ vs. effective normal load $f^{\n}_{\s}$ for the rail system with different rail inclination  ranging from $0\degree$, $25\degree$, $45\degree$ and $60\degree$. All data sets can be fitted with a simple linear relation \eqref{eq:muroto}.}
	\label{fig:rail_1}
\end{figure}

The surface roughness of the rail slopes cannot be ignored and has a significant effect on our measurements. According to the correlation between rolling resistance and surface roughness\new{, as discussed in section \ref{sec:45}, roughnesses between 2 and 5\,nm lead to measurably smaller lateral forces, see \figref{fig:rail_1}. The difficulty lies in preparing comparable rail systems with a constant low surface roughness smaller than 1\,nm for all rail inclinations, which would allow us to focus on the effect of normal loads.} Until now, this is not realized and all presented values for the different rail systems have a systematic error due to variable surface roughness.

\read{Fig. \ref{fig:rail_1} shows the effective lateral forces, $f_\s^\l$, and normal loads, $f_\s^\n$, at the substrate contacts, which are calculated from the  measured lateral force, $f_\i^\l$,  and the applied normal load, $f_\i^\n$, according to \eqref{eq:forces4}.}

All results follow linear trends, confirming a modified Coulomb law, but the slopes, which are the effective friction coefficients $\mu^\rail$, show a clear correlation with increasing  inclination  of the rail, \new{which will be further analysed in section \ref{sec:48}}. Effects of sliding are excluded, since a pure sliding test (nano\-indenter colloid probe) on a $25\degree$ rail with a normal load of $100\,\mathrm{\mu N}$ results in an average lateral force of $26\,\mathrm{\mu N}$ $\pm 1.71\,\mathrm{\mu N}$. This value is more than one order of magnitude higher compared to the measured lateral force with a single free rolling microsphere inside the $25\degree$ rail at corresponding normal load. 

It is expected that for higher inclination the measured lateral force increases \new{due to the larger}  torsion component. 
\read{The intercepts of the trend lines with the horizontal axis, taken as the values of rail adhesion, $h^\rail$, are lower than for pure rolling. This suggests that adhesion plays a less dominant role during the combines rolling/torsion motion than for pure rolling on a plane surface.}

Even though the experimental results presented up to now provide evidence of an increased effect of torsion with increasing  inclination, a deeper insight into appropriate contact models is necessary to understand the evolution of the effective friction coefficient, $\mu^\rail$, with respect to  inclination. In the following section, focusing on contact models and simulation, among other considerations, we strive to quantify the impact of rolling and torsional friction seen here.

\subsection{Analysis of combined rolling and torsion}\label{sec:48}

\renewcommand{\t}{\tan\theta}
We now consider the case where the bottom wall consists of a v-shaped rail with inclination angle $\theta$, see \figref{fig:forces0}c. Thus we have three contact points denoted by the superscripts $\i$, $\s$ and $\s'$. Only the contact points at the substrate have nonzero torsion velocity and thus nonzero torsion resistance.

We assume that the forces and torques acting at the two contact points with the substrate are equal in magnitude
\begin{subequations}
\begin{gather}
|\vec f^{\n}_{\s}| = |\vec f^{\n}_{s'}|,~ 
 |\vec f^{\l}_{\s}| = |\vec f^{\l}_{s'}|,\\
 |\vec M^{\ro}_{\s}| = |\vec M^{\ro}_{s'}|,\mbox{ }
 |\vec M^{\to}_{\s}| = |\vec M^{\to}_{s'}|.\label{eq:forces3b}\end{gather}
 \end{subequations}
For steady motion, the $x$ and $y$-components of the force balance yield
\[ |\vec f^{\l}_{\i}| = 2|\vec f^{\l}_{\s}| \mbox{ and } |\vec f^{\n}_{\i}| = 2\cos\theta |\vec f^{\n}_{\s}|, \label{eq:forces4}\]
and the torque balance in $z$-direction yields
\begin{gather} 
r_{\i}|\vec f^{\l}_{\i}|+2r_{\s} \cos\theta |\vec f^{\l}_{\s}|\nonumber\\
=|\vec M^{\ro}_{\i}| + 2\cos\theta |\vec M^{\ro}_{\s}| + 2\sin\theta |\vec M^{\to}_{\s}|.
\label{eq:forces6}
\end{gather}

\begin{figure}[tbp]
\centering
\includegraphics[width=1\columnwidth]{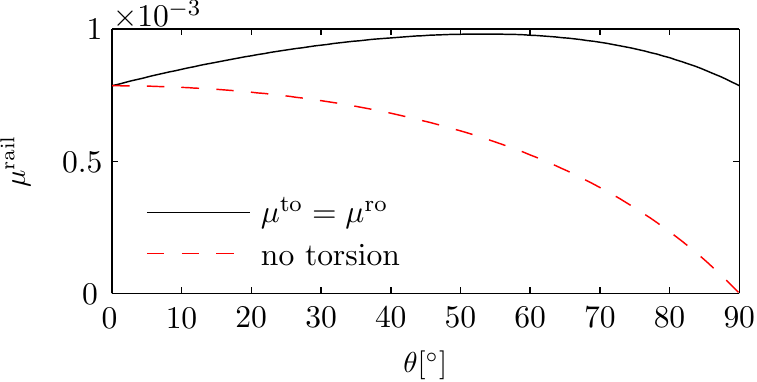}
\caption{Combined rolling/ torsion friction coefficient, see \eqref{eq:muroto}, plotted against rail inclination for rails with constant RMS roughness $\Delta=0.3\,$nm, for the torsionless case ($\mu^{\to}=0$, dashed line) and equal rolling/ torsion coefficients ($\mu^{\to}=\mu^{\ro}$, solid line). 
}
\label{fig:rail_3}
\end{figure}

\read{
We assume that the particle does not slide at the indenter contact, since the lateral forces measured in \figref{fig:rail_1} are too small to be attributed to sliding friction. 
Thus, we conclude that both rolling and torsion torques reach their threshold value.
Substituting \eqref{eq:forces4} and \eqref{eq:yield} into \eqref{eq:forces6} 
yields relations between measured normal and lateral forces - or, as we chose here, between normal and lateral forces at the substrate,
\begin{align}
2(r_\i+r_\s\cos\theta) |\vec f^{\l}_{\s}| &=\mu_{\i}^{\ro}\la_{\i}^{\ro}(2\cos\theta f^{\n}_{\s}+\h_{\i}^{\ro})\nonumber\\
&+2\cos\theta \mu_{\s}^{\ro}\la_{\s}^{\ro}(f^{\n}_{\s}+\h_{\s}^{\ro})\nonumber\\
&+2\sin\theta \mu_{\s}^{\to}\la_{\s}^{\to}(f^{\n}_{\s}+\h^{\to}_\s).\label{eq:torsion}
\end{align}
In short, the relation plotted in \figref{fig:rail_1} satisfies%
\begin{subequations}
\begin{align}
|\vec f^{\l}_{\s}| = \mu^{\rail}(f^{\n}_{\s}+h^\rail), \label{eq:muroto}
\end{align}
where
\begin{align}
&\mu^{\rail} \!= \frac{2\cos\theta\bar\mu^{\ro} \bar\la^{\ro}+\sin\theta \mu_{\s}^{\to} \la_{\s}^{\to}}{r_\i+r_\s\cos\theta},\label{eq:murail}\\
&h^\rail \!= \frac{\frac{1}{2}\mu_{\i}^{\ro}\la_{\i}^{\ro} \h_{\i}^{\ro}\!+\!\cos\theta \mu_{\s}^{\ro}\la_{\s}^{\ro}\h_{\s}^{\ro}
\!+\!\sin\theta \mu_{\s}^{\to}\la_{\s}^{\to} \h^{\to}_\s}{\mu^{\rail}(r_\i+r_\s\cos\theta)}.
\end{align}
\end{subequations}
}%
\read{Thus, the measured lateral force is due to pure rolling for zero inclination, and due to pure torsion in the limit case of vertical rails, $\theta\TO90^\circ$.}
For intermediate angles, both  torsion and rolling friction contribute to the lateral force. The contribution of rolling friction to the effective rolling{/}torsion friction coefficient $\mu^{\rail}$ decreases for higher inclination, while the effect of torsion friction increases, if the surface properties of the rails would be independent of inclination, as illustrated in \figref{fig:rail_3}.

\begin{figure}[tbp]
	\centering
		\includegraphics[width=1\columnwidth]{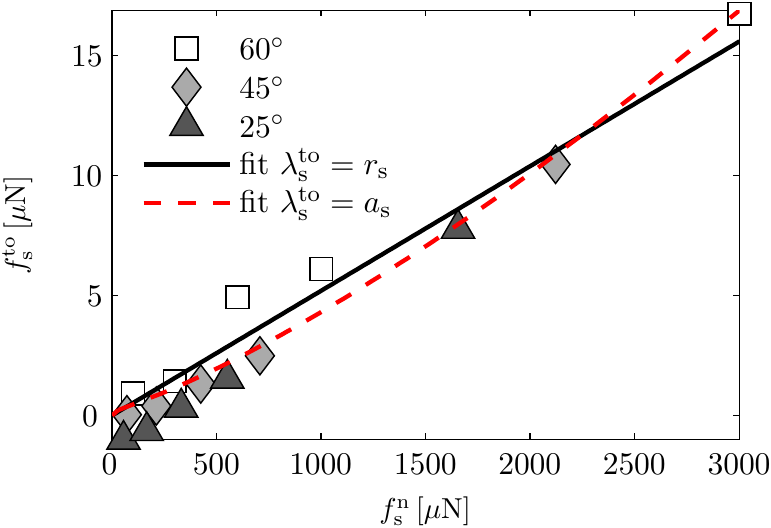}
		\caption{Torsion contribution to the lateral force $f_{\s}^{\to}$, see \eqref{eq:muto2}, vs. effective normal load $f^{\n}_{\s}$ for rail system friction measurements with different rail  inclinations $25\degree$, $45\degree$ and $60\degree$. All data sets are fitted to a single relation; lines show fits of $\mu_{\s}^{\to}$ for $\la_{\s}^{\to}=r_{\s}$ (solid line) and $\la_{\s}^{\to}=a_{\s}$ (dashed red line).}
	\label{fig:rail_2}
\end{figure}

\read{To measure the torsional component of the lateral force, the influence of the rolling friction is subtracted from the measured lateral force,
\begin{align}
f_{\s}^{\to}&=\big[2(r_\i+r_\s\cos\theta)|\vec{f}^\l_\s|-4\cos\theta\bar\mu^\ro\bar\la^\ro f^\n_\s\label{eq:muto2}\\
&-(\mu_{\i}^{\ro}\la_{\i}^{\ro}\h_{\i}^{\ro}+2\cos\theta\mu_{\s}^{\ro}\la_{\s}^{\ro}\h_{\s}^{\ro})\big]/(2r_\s\sin\theta)\nonumber
.
\end{align}
Note that substituting \eqref{eq:torsion} into \eqref{eq:muto2} and using Eqs. \eqref{eq:barlaro} and \eqref{eq:forces4}  yields
\begin{align}
f_{\s}^{\to}= \mu_{\s}^{\to} \frac{\la_{\s}^{\to}}{r_\s} (f^{\n}_{\s}+\h^{\to}_\s),\label{eq:muto}
\end{align}
which resembles \eqref{eq:fs1b} in shape.

\begin{figure}[tbh]
\centering
\includegraphics[width=1\columnwidth]{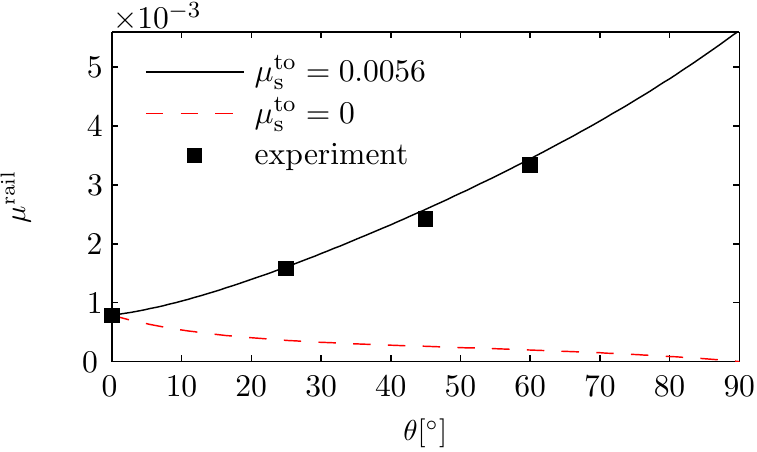}
\caption{Combined rolling/torsion friction coefficient, see \eqref{eq:muroto}, for the rail experiment shown in \figref{fig:forces0}c. To observe the effect of torsion, simulational/analytical results \new{(including the effect of roughness, see \figref{fig:rolling_2})} are shown for both $\mu_{\s}^{\to}=0$ (red dashed line) and $\mu_{\s}^{\to}=0.0056$ (solid line) with $\la_{\s}^{\to}=r_{\s}^{\to}$.}
\label{fig:rail}
\end{figure}

Eqn. \eqref{eq:muto2} cannot be evaluated exactly, since the rolling friction factor and adhesion are only measured as averaged values, see \eqref{eq:barto}.  Therefore, to evaluate \eqref{eq:muto2}, we assume that 
$\mu_{\i}^{\ro}\la_{\i}^{\ro}\h_{\i}^{\ro} +2\cos\theta$ $\mu_{\s}^{\ro}\la^{\ro}_\s\h_{\s}^{\ro}=(1+2\cos\theta)\bar\mu^{\ro}\bar\la^{\ro}\bar\h^{\ro}$, which is exactly satisfied if the rolling friction factor and rolling adhesion force at the substrate and at the indenter would be equal. The force $f_\s^\to$ is plotted in \figref{fig:rail_2}. 
Note that, while the last assumption is not generally justified for all combinations of indenter and substrate surfaces, it only influences the fit of the torsion adhesion force (which we do not discuss further here), but has no influence on the torsion friction coefficient. For the special case of $\theta=60^\circ$, however, the assumption can be proven using \eqref{eq:barto}.
Further, the average rolling friction coefficient $\bar\mu^\ro$ and rolling adhesion force $\bar\h^\ro$ in \eqref{eq:muto2} is a function of the rail roughness $\Delta$, as shown in \figref{fig:rolling_2}. 
}

The torsion friction shows an almost constant slope, which can be fitted by
\[\mu_{\s}^{\to} =0.0056\mbox{ for }\la_\s^{\to}=r_\s.\] and is shown in \figref{fig:rail_2}. 
The torsion adhesion force, $\h^{\to}_\s$, cannot be defined properly, as the measurement error for small normal forces is too large.
\new{This fit is then used to obtain analytical predictions for the rail friction coefficient that are in good agreement with the experimental measurements, as shown in Figure \ref{fig:rail}, where the rail roughness is fitted according to \eqref{eq:rofit}.}  

The measurements were also fitted by assuming a scaling with the contact radius, resulting in 
\[\mu_{\s}^{\to} =0.064\mbox{ for }\la_\s^{\to}=a_\s.\]
Both fits agree well with the data, \new{as  shown in  \figref{fig:rail_2}.} Thus, the measurements are not able to  \new{distinguish if the torsion friction factor is independent of the normal force or scales with the (normal-force dependent) contact radius}. Further, the sliding friction coefficient for the given rail roughness can be extrapolated from the data in \tabref{tab:surf} to be in the narrow interval $0.63\leq \mu_{\s}^{\l}\leq 0.72$. Thus,  \new{the dependence of $\mu_{\s}^{\to}$ on the sliding friction coefficient $\mu_\s^\l$, as predicted by \eqref{eq:muto_to},} is \new{in disagreement} with the given data. Instead, the measured coefficient is about seven times smaller than the predicted value of $\mu_{\s}^{\to}=(2/3)\mu^{\l}\mbox{ for }\la_\alpha^{\to}=a_\alpha$, unless a smaller effective radius is used. A more detailed study with different particle radii is necessary to find correlations with either particle or effective contact radii.

\section{Summary and Conclusions}
The measurement of different motions such as sliding, rolling and torsion of individual very small particles was made
possible by a relatively simple nanoindenter setup, used as a platform where experiment and model/theory meet.
Compared with other techniques used in literature,
a nanoindenter featuring the option to measure not only normal loads but also lateral forces
was used. Even though the interpretation/evaluation of the results still leaves various open questions,
a big step towards understanding particle interactions and to directly obtain contact model
parameters was made.

\subsection{Normal force}
Experimentally, the normal force between  ${\sim}10\mu$m borosilicate spheres and different rough Si surfaces is well described by
a Hertzian law but the adhesion force decays with increasing surface roughness (on the scale of $0.3-3$nm). 
The elastic Hertzian contact model approximates well the radius $a$ of the contact area.
While the effect of the considerable surface roughness on the normal force is surprisingly small, it should lead
to notable plastic deformations of the contacting micro-asperities.

\subsection{Sliding, Rolling and Torsion}
Next, the influence of surface roughness and adhesion on pure sliding and pure rolling measurements
is studied for the same particles on Si surfaces. In addition, a series of experiments in a rail features a
combination of torsion and rolling motion, for which a model to distinguish between rolling and
torsion friction is presented.
\read{Sliding and torsion occurs only at the contact with the substrate; in contrast, rolling torques
occur at both substrate and indenter contact, thus being an average quantity from two contacts.}
All 
motion types (pure sliding, pure rolling
and mixed torsion and rolling) are represented as relations between the measured lateral force,
the applied normal load, via an effective friction coefficient $\mu$, and an effective adhesion $h$:%
\begin{equation}
f^{\rm l} = \mu ( f^{\rm n} + h ) ~,
\end{equation}
where typically the smallest forces are either unreliable or display a mix of different motion
degrees of freedom, while the largest normal loads reach levels of $3000\,\mu$N.

All the data can be fitted by either a linear or a non-linear model. The former assumes independence of adhesion and friction coefficient with respect to the normal force. This model leads to constant friction factors for rolling and torsion (when scaled with the particle radius) that are much smal\-ler than for sliding, but also to highly different predictions for the effective adhesion for the different motions, being highest for pure rolling.
The latter non-linear model \red{still assumes independence of adhesion with respect to the normal force, but} takes into account the dependence of the \red{rolling and torsion friction factors} on the (predicted) contact radius by introducing it as a length scale into the yield criteria for the rolling and torsion torques.

\subsection{Linear vs. Non-Linear Model}
More specifically, for the linear model we conclude the following:
In the (pure) sliding case, an increase of the friction coefficient with increasing RMS roughness is observed,
which we attribute to the considerable increase of lateral plastic deformations of the higher asperities in the contact area.
The effective sliding adhesion is about a factor of 4 larger than the normal adhesion.
In contrast, for pure rolling the friction coefficient seems to decrease with increasing surface roughness,
while the effective adhesion enormously increases by 2.5 orders of magnitude.
This indicates that rolling, in our situation, is either mostly adhesion dominated, or that the linear model is inadequate.
Some random deviations perpendicular to the motion of the indenter were observed, but the effect was
considered minor.
\redd{The non-linear rolling friction model produces qualitatively similar results; thus the measurements do not allow us to conclude whe\-ther the friction factor scales with the contact radius and thus with the normal force.}
%

In the mixed, rail configuration, we observe that the effective friction increases with higher rail inclination angle.
The details of the preparation of the rail-structured silicon surfaces with different angles introduces a new
source of (uncontrollable) surface roughness,
which makes it challenging to compare the different rail inclination angle data - especially
since their interpretation relies on pure rolling results that come from a different surface:
Kinematic arguments allow us to quantify the torsion contribution to the effective friction of the rail set-up,
which increases with higher rail inclination, by separating the torsion from the mixed rolling and torsion contributions.
The extracted torsion friction coefficient -- for the first time to our knowledge -- shows \emph{little dependence} on the
surface roughness at different angles.
Unfortunately, in this framework nothing can be said about the effective torsion adhesion.

%

Taking the RMS roughness results into account, \new{the torsion friction factor,
$\mu_{\s}^{\to}\la_\s^\to$, was determined to be $\approx 0.0056 r_\alpha$ by the linear fit} (assuming that the torsion resistance scales with the particle radius), \new{and $\approx 0.065 a_\alpha$ by the non-linear fit} (assuming that the torsion resistance scales with the contact radius). Both fits agree well with the data in the available range; \red{thus, as in the case of rolling friction,}
\redd{the present set of data does not allow to distinguish which of the two models is the true one.}

\subsection{Outlook}
From the experiments presented here, no conclusion can be drawn about
dependencies on the particle radius, since the particle size was kept constant.
Further experiments with varying particle radii, e.g. 2.5-$50\,\mu$m, are in
progress and will allow to determine the dependencies on the particle radius.
This, however, poses new challenges, as surface roughness and non-sphericity vary
with particle size too.

Further, a deeper understanding of the effect of surface roughness and adhesion on the
frictional behavior is required. Experimentally, the dependence on the contact adhesion,
due to liquid bridges and van-der-Waals forces, should be studied by varying the relative
humidity and the hydrophobicity of the surfaces. The experimental results can then be used
to calibrate an adhesive elasto-plastic force model applied to a continuum model of the
particle and the contacting (rough) surfaces.

In the future, our experimental results can be used as starting input parameters for
advanced contact models in many-particle simulations to predict the bulk behavior of
fine powders, where adhesion affects all the degrees of freedom for particle-particle contacts,
forces, and torques.

\section*{Acknowledgments}
The authors would like to thank \new{Lothar Brendel for many helpful discussions}. 
We also thank the German Research Foundation (DFG) for financial support. 
This work is carried out within the framework of the Key Research Program (SPP 1486 PiKo \quotes{Particles in Contact}) grants LU 450/10-1, LU 450/10-2, STA 1021/1-1 and STA 1021/1-2.
The numerical solutions of the contact models in this paper were \new{carried out} using the 
open-source code MercuryDPM (\texttt{mercurydpm.org}). 

\bibliographystyle{unsrt}

\begin{thebibliography}{10}

\bibitem{Ducker1991}
W.~A. Ducker, T.~J. Senden, and R.~M. Pashley.
\newblock Direct measurement of colloidal forces using an atomic force
  microscope.
\newblock {\em Nature}, 353(6341):239--241, 1991.

\bibitem{Butt1991}
H.~J. Butt.
\newblock Measuring electrostatic, van der {W}aals, and hydration forces in
  electrolyte-solutions with an atomic force microscope.
\newblock {\em Biophysical Journal}, 60(6):1438--1444, 1991.

\bibitem{Sitti2000}
M.~Sitti and H.~Hashimoto.
\newblock Controlled pushing of nanoparticles: Modeling and experiments.
\newblock {\em {IEEE-ASME} Transactions on Mechatronics}, 5(2):199--211, 2000.

\bibitem{Sitti2004}
M.~Sitti.
\newblock Atomic force microscope probe based controlled pushing for
  nanotribological characterization.
\newblock {\em {IEEE-ASME} Transactions on Mechatronics}, 9(2):343--349, 2004.

\bibitem{Sumer2008}
B.~Sumer and M.~Sitti.
\newblock Rolling and spinning friction characterization of fine particles
  using lateral force microscopy based contact pushing.
\newblock {\em Journal of Adhesion Science and Technology}, 22(5-6):481--506,
  2008.

\bibitem{Liu2007}
D.~L. Liu, J.~Martin, and N.~A. Burnham.
\newblock Optimal roughness for minimal adhesion.
\newblock {\em Applied Physics Letters}, 91(4), 2007.

\bibitem{Rabinovich2000a}
Y.~I. Rabinovich, J.~J. Adler, A.~Ata, R.~K. Singh, and B.~M. Moudgil.
\newblock Adhesion between nanoscale rough surfaces - i. role of asperity
  geometry.
\newblock {\em Journal of Colloid and Interface Science}, 232(1):10--16, 2000.

\bibitem{Rabinovich2000b}
Y.~I. Rabinovich, J.~J. Adler, A.~Ata, R.~K. Singh, and B.~M. Moudgil.
\newblock Adhesion between nanoscale rough surfaces - ii. measurement and
  comparison with theory.
\newblock {\em Journal of Colloid and Interface Science}, 232(1):17--24, 2000.

\bibitem{Matope2012}
S.~Matope, Y.~I. Rabinovich, and A.~F. Van~der Merwe.
\newblock Van der {W}aals interactions between silica spheres and metallic thin
  films created by e-beam evaporation.
\newblock {\em Colloids and Surfaces a-Physicochemical and Engineering
  Aspects}, 411:87--93, 2012.

\bibitem{Korayem2011}
M.~H. Korayem and M.~Zakeri.
\newblock Dynamic modeling of manipulation of micro/nanoparticles on rough
  surfaces.
\newblock {\em Applied Surface Science}, 257(15):6503--6513, 2011.

\bibitem{Saito2002}
S.~Saito, H.~T. Miyazaki, T.~Sato, and K.~Takahashi.
\newblock Kinematics of mechanical and adhesional micromanipulation under a
  scanning electron microscope.
\newblock {\em Journal of Applied Physics}, 92(9):5140--5149, 2002.

\bibitem{Peri2008}
M.~D.~Murthy Peri and Cetin Cetinkaya.
\newblock Adhesion characterization based on rolling resistance of individual
  microspheres on substrates: Review of recent experimental progress.
\newblock {\em Journal of Adhesion Science and Technology}, 22(5-6):507--528,
  2008.

\bibitem{Ding2007}
W.~Ding, A.~J. Howard, M.~D.~Murthy Peri, and C.~Cetinkaya.
\newblock Rolling resistance moment of microspheres on surfaces: contact
  measurements.
\newblock {\em Philosophical Magazine}, 87(36):5685--5696, 2007.

\bibitem{Vilt2011}
Steven~G. Vilt, Nathaniel Martin, Clare McCabe, and G.~Kane Jennings.
\newblock Frictional performance of silica microspheres.
\newblock {\em Tribology International}, 44(2):180--186, 2011.

\bibitem{hertz1882b}
H.~Hertz.
\newblock {\"U}ber die {B}er\"uhrung fester elastischer {K}\"orper.
\newblock {\em J. f\"ur die reine u. angew. Math.}, 92, 1882.

\bibitem{Luding2008c}
S.~Luding.
\newblock Cohesive, frictional powders: contact models for tension.
\newblock {\em Granular Matter}, 10(4):235--246, 2008.

\bibitem{Tomas20075925}
J.~Tomas.
\newblock Adhesion of ultrafine particles: {E}nergy absorption at contact.
\newblock {\em Chemical Engineering Science}, 62(21):5925 -- 5939, 2007.

\bibitem{SinghMagnanimoLuding2013}
A.~Singh, V.~Magnanimo, and S.~Luding.
\newblock Mesoscale contact models for sticky particles.
\newblock {\em Powder Technology}, 2013 (submitted).

\bibitem{SunZengYu2013}
W.~Sun, Q.~Zeng, and A.~Yu.
\newblock Calculation of noncontact forces between silica nanospheres.
\newblock {\em Langmuir}, 29(7):2175--2184, 2013.

\bibitem{ShojaaeeBrendelTorokWolf2012}
Z.~Shojaaee, L.~Brendel, J.~T\"or\"ok, and D.~E. Wolf.
\newblock Shear flow of dense granular materials near smooth walls. ii. block
  formation and suppression of slip by rolling friction.
\newblock {\em Phys. Rev. E}, 86:011302, 2012.

\bibitem{WangZhuLudingYu2013}
X.~Wang, H.~P. Zhu, S.~Luding, and A.~B. Yu.
\newblock Regime transitions of granular flow in a shear cell: A
  micromechanical study.
\newblock {\em Phys. Rev. E}, 88:032203, 2013.

\bibitem{morita1989control}
M~Morita, T~Ohmi, E~Hasegawa, M~Kawakami, and K~Suma.
\newblock Control factor of native oxide growth on silicon in air or in
  ultrapure water.
\newblock {\em Applied Physics Letters}, 55(6):562--564, 1989.

\bibitem{vanZwol2008}
P.~J. van Zwol, G.~Palasantzas, M.~van~de Schootbrugge, J.~Th.~M. de~Hosson,
  and V.~S.~J. Craig.
\newblock Roughness of microspheres for force measurements.
\newblock {\em Langmuir}, 24(14):7528--7531, 2008.

\bibitem{FuchsMeyerStaedlerJiang2013}
R.~Fuchs, J.~Meyer, T.~Staedler, and X.~Jiang.
\newblock Sliding and rolling of individual micrometre sized glass particles on
  rough silicon surfaces.
\newblock {\em Tribology{--}Materials, Surfaces \& Interfaces}, 7(2):103--107,
  2013.

\bibitem{ButtJaschke1995}
H.-J. Butt and M.~Jaschke.
\newblock Calculation of thermal noise in atomic force microscopy.
\newblock {\em Nanotechnology}, 6(1):1, 1995.

\bibitem{kuhn2004contact}
M.R. Kuhn and K.~Bagi.
\newblock Contact rolling and deformation in granular media.
\newblock {\em International Journal of Solids and Structures},
  41(21):5793--5820, 2004.

\bibitem{tykhoniuk2007ultrafine}
R.~Tykhoniuk, J.~Tomas, S.~Luding, M.~Kappl, L.~Heim, and H.-J. Butt.
\newblock Ultrafine cohesive powders: from interparticle contacts to continuum
  behaviour.
\newblock {\em Chemical Engineering Science}, 62(11):2843--2864, 2007.

\bibitem{brilliantov1998rolling}
N.~V. Brilliantov and T.~P{\"o}schel.
\newblock Rolling friction of a viscous sphere on a hard plane.
\newblock {\em EPL (Europhysics Letters)}, 42(5):511, 1998.

\bibitem{luding1998collisions}
S.~Luding.
\newblock Collisions \& contacts between two particles.
\newblock {\em NATO ASI Series E Applied Sciences-Advanced Study Institute},
  350:285--304, 1998.

\bibitem{kuwabara1987restitution}
G.~Kuwabara and K.~Kono.
\newblock Restitution coefficient in a collision between two spheres.
\newblock {\em Jpn. J. Appl. Phys. 1}, 26(8):1230--1233, 1987.

\bibitem{thornton2012investigation}
C.~Thornton, S.~J. Cummins, and P.~W. Cleary.
\newblock An investigation of the comparative behaviour of alternative contact
  force models during inelastic collisions.
\newblock {\em Powder Technology}, 233(0):30 -- 46, 2013.

\bibitem{Luding1998}
Stefan Luding.
\newblock Collisions \& contacts between two particles.
\newblock {\em NATO ASI Series E Applied Sciences-Advanced Study Institute},
  350:285--304, 1998.

\bibitem{mindlin1949compliance}
R.D. Mindlin.
\newblock Compliance of elastic bodies in contact.
\newblock {\em J. of Appl. Mech.}, 16, 1949.

\bibitem{mindlin1953elastic}
R.D. Mindlin and H.~Deresiewicz.
\newblock Elastic spheres in contact under varying oblique forces.
\newblock {\em J. of Appl. Mech.}, 20, 1953.

\bibitem{BrendelTorokKirschBrockel2011}
L.~Brendel, J.~T\"or\"ok, R.~Kirsch, and U.~Br\"ockel.
\newblock A contact model for the yielding of caked granular materials.
\newblock {\em Granular Matter}, 13(6):777--786, 2011.

\bibitem{yang2000computer}
R.~Y. Yang, R.~P. Zou, and A.~B. Yu.
\newblock Computer simulation of the packing of fine particles.
\newblock {\em Physical Review E}, 62(3):3900, 2000.

\bibitem{beer2003vector}
F.~P. Beer, J.~E. Russell~Johnston, E.R.~Jr. Johnston, E.R. Eisenberg, W.E.
  Clausen, and G.H. Staab.
\newblock {\em Vector Mechanics for Engineers: Statics and Dynamics}.
\newblock McGraw-Hill, 2003.

\bibitem{BrilliantovPoschel1998}
N.~V. Brilliantov and T.~P\"oschel.
\newblock Rolling friction of a viscous sphere on a hard plane.
\newblock {\em EPL (Europhysics Letters)}, 42(5):511, 1998.

\bibitem{johnson1984contact}
K.~L. Johnson.
\newblock {\em Contact mechanics}.
\newblock Cambridge University Press, Cambridge, UK, 1984.

\bibitem{deresiewicz1954contact}
H.~Deresiewicz.
\newblock Contact of elastic spheres under an oscillating torsional couple.
\newblock {\em Journal of Applied Mechanics}, 21:52--56, 1954.

\bibitem{dintwa2005torsion}
E.~Dintwa, M.~van Zeebroeck, E.~Tijskens, and H.~Ramon.
\newblock Torsion of viscoelastic spheres in contact.
\newblock {\em Granular Matter}, 7(2):169--179, 2005.

\bibitem{Farkas2003}
Z.~Farkas, G.~Bartels, T.~Unger, and D.~E. Wolf.
\newblock Frictional coupling between sliding and spinning motion.
\newblock {\em Physical Review Letters}, 90(24), 2003.

\bibitem{Tabor1996}
D.~Tabor.
\newblock Indentation hardness: Fifty years on - a personal view.
\newblock {\em Philosophical Magazine a-Physics of Condensed Matter Structure
  Defects and Mechanical Properties}, 74(5):1207--1212, 1996.

\bibitem{Bhushan1998}
B.~Bhushan.
\newblock Contact mechanics of rough surfaces in tribology: multiple asperity
  contact.
\newblock {\em Tribology Letters}, 4(1):1--35, 1998.

\bibitem{Lothar}
Lothar Brendel.
\newblock private communication.

\bibitem{Briggs1976}
G.~A.~D. Briggs and B.~J. Briscoe.
\newblock Effect of surface-roughness on rolling friction and adhesion between
  elastic solids.
\newblock {\em Nature}, 260(5549):313--315, 1976.

\end{thebibliography}

\end{document}